\documentclass[10pt]{article}

\usepackage[all,cmtip]{xy}
\usepackage{framed}
\usepackage{hyperref}
\usepackage{amsmath,amssymb,bbm}

\begin{document}

\rightline{LTH 1063} 

\vskip 1.5 true cm  
\begin{center}  
{\Large \textbf{
Hessian geometry and the holomorphic anomaly
}}\\[1em]



\vskip 1.0 true cm   
{G.~L. Cardoso $^1$, T.~Mohaupt $^2$} \\[1em] 
$^1${Center for Mathematical Analysis, Geometry and Dynamical Systems\\
  Instituto Superior T\'ecnico\\ Universidade de Lisboa\\
  Av. Rovisco Pais\\ 1049-001 Lisboa, Portugal\\
  gcardoso@math.ist.ut.pt
}\\[1em]  
$^2${Department of Mathematical Sciences\\ 
University of Liverpool\\
Peach Street \\
Liverpool L69 7ZL, UK\\  
thomas.mohaupt@liv.ac.uk}\\[1em]

November 20, 2015. Revised: February 12, 2016.

\end{center}  
\vskip 1.0 true cm

\begin{abstract}  

\noindent  
We present a geometrical framework which incorporates higher
derivative corrections to the action of ${N}=2$
vector multiplets in terms of an enlarged scalar manifold which
includes a complex deformation parameter. This enlarged space
carries a deformed version of special K\"ahler geometry which 
we characterise. The holomorphic anomaly equation arises in this
framework from the integrability condition for the existence of
a Hesse potential. 
\end{abstract}  
\newpage

\section{Introduction \label{intro}}

Quantum gravity is expected to manifest itself in an effective field
theory framework through higher derivative terms. Supersymmetry provides
some control over such terms, and in a theory of ${N}=2$ vector
multiplets coupled to supergravity a certain class of higher derivative
terms can be described by generalizing the prepotential which encodes
all the couplings at the two-derivative level. While this function remains
holomorphic within a Wilsonian framework, the inclusion of threshold
corrections due to massless particles is known to induce non-holomorphic
corrections to the couplings. In fact these corrections are required
for consistency with electric-magnetic duality and are essential for incorporating
higher derivative corrections to black hole entropy. While supergravity
provides a powerful tool to organise an effective action for quantum gravity,
the actual computation of couplings requires a specific 
theory.
String theory is the natural candidate, 
and in particular the higher derivative corrections to ${N}=2$
vector multiplets are captured by the topological string. However,
the relation between supergravity and the topological string is subtle,
and non-holomorphic corrections are incorporated differently in the
respective formalisms. In this paper we develop a new geometrical 
description of the higher derivative corrections on the supergravity 
side, by showing that they can be understood in terms of an extended
scalar manifold which carries a deformed version of special geometry.
We also derive various exact relations between
the variables used in supergravity and in the topological string. 
The most interesting result we obtain is that the holomorphic
anomaly equation which controls the non-holomorphic corrections in 
both the supergravity and topological string formalism can be derived
from the integrability condition for the existence of a Hesse potential
on the extended scalar manifold.

Let us next introduce our topic in more technical terms. 
In four dimensions, the complex scalar fields residing in $N=2$ vector multiplets 
parametrize a scalar manifold which
is the target space of the non-linear sigma-model that enters
in the Wilsonian Lagrangian describing the couplings of $N=2$ vector multiplets
at the two-derivative level. The scalar manifold is
an affine special K\"ahler manifold in global
supersymmetry, and a projective special K\"ahler manifold in local 
supersymmetry; both types of target space geometry are referred to as
special geometry
 \cite{deWit:1984pk,deWit:1984px,Strominger:1990pd,Castellani:1990tp,Castellani:1990zd,
Freed:1997dp,2002JGP....42...85A}. 
Special geometry, when formulated
in terms of complex variables $Y^I$, is encoded in a holomorphic function $F^{(0)} (Y)$,
called the prepotential. 
When formulated in terms of special real coordinates, it is the Hesse potential that plays a central role.
For affine special K\"ahler manifolds, the Hesse potential is related to the prepotential by a Legendre transform
\cite{Cortes:2001}.

When coupling the $N=2$ vector multiplets to the square of the Weyl multiplet, the resulting Wilsonian
Lagrangian, which now contains higher derivative terms proportional to the square of the Weyl tensor,
in encoded in a generalized prepotential $F(Y, \Upsilon)$, where $\Upsilon$ denotes a complex
scalar field residing in the lowest component of the square of the Weyl multiplet. 
The complex scalar fields $(Y^I, \Upsilon)$ will be called supergravity variables in the following.
The prepotential $F^{(0)}(Y)$ is obtained from $F(Y, \Upsilon)$ by setting $\Upsilon =0$.
Electric-magnetic duality, a central feature of $N=2$ systems based on vector multiplets, 
then acts by symplectic transformations of the vector $(Y^I, F_I)$, where $F_I = \partial F/\partial Y^I$.
While $F(Y, \Upsilon)$ itself does not transform as a function under symplectic transformations, $F_{\Upsilon} = 
\partial F/\partial \Upsilon$ does \cite{deWit:1996ix}. The associated Hesse potential, obtained by a Legendre
transform of ${\rm Im} \, F(Y, \Upsilon)$, is also a symplectic function.

Away from the Wilsonian limit, the coupling functions encoded in $F_{\Upsilon}$
receive non-holomorphic corrections, in general.
In supergravity models arising from string theory, these modified coupling functions can be derived
in the context of topological string theory \cite{Bershadsky:1993cx,Antoniadis:1993ze}. The precise relation between these two computations is
subtle, however \cite{Cardoso:2014kwa}. The coupling functions computed in topological string theory
depend on stringy variables $({\cal Y}^I, \Upsilon)$ that do not coincide with the supergravity variables
$(Y^I, \Upsilon)$ (unless $\Upsilon =0$). 
The precise relation between these two sets of coordinates
was discussed in \cite{Cardoso:2014kwa} and was used to express the supergravity Hesse potential (which is a symplectic function)  in terms
of stringy variables. The Hesse potential is not any longer obtained from  the holomorphic generalized
prepotential $F(Y, \Upsilon)$ that characterizes the Wilsonian Lagrangian.  Instead, it is computed from a deformed version of $F$ that is not any longer
holomorphic.  It was then laboriously shown, by means of power law expansions, that the Hesse potential
contains a unique subsector ${\cal H}^{(1)}$ that comprises coupling functions $F^{(n)}({\cal Y}, \bar{\cal Y})$
that, for $n \geq 2$, satisfy the holomorphic anomaly equation of topological string theory, 
\begin{equation}
\frac{\partial F^{(n)}}{\partial \bar{\cal Y}^I} =  
i \bar{F}^{(0)JK}_I \left( \sum_{r=1}^{n-1} \partial_J F^{(r)} 
\partial_K F^{(n-r)} - 2 \alpha D_J \partial_K F^{(n-1)} 
\right) \;.
\label{anomal}
\end{equation}
Here, $\alpha$ denotes the deformation parameter that characterizes the 
deviation from the Wilsonian limit. The superscript $(0)$ in
$\bar{F}^{(0)JK}_I$ indicates that this quantity has been formed by
taking derivatives of $F^{(0)}(Y)$, and that indices have been 
raised using the inverse $N_{(0)}^{IJ}$
corresponding metric $N^{(0)}_{IJ} = 2 \mbox{Im} F^{(0)}_{IJ}$. If no such 
superscript is present, as in $\bar{F}^{JK}_I$, then it is understood
that we take derivatives of the generalized prepotential $F(Y,\Upsilon)$, 
and that indices are lowered and raised using $N_{IJ}= 2 \mbox{Im} F_{IJ}$. 
This convention is applied throughout the paper.

While in topological string theory $\alpha=-\frac{1}{2}$
\cite{Bershadsky:1993cx}, one may ask a more general question, namely whether irrespective of the value of $\alpha$, the holomorphic anomaly equation \eqref{anomal}
can be understood in terms of Hessian structures and Hessian geometry. This is indeed the case, as we will
show in this paper. Namely, the anomaly equation \eqref{anomal} may be viewed as the integrability condition 
for the existence of a Hesse potential in supergravity. This is simplest to establish in the case when $\alpha =0$, as we will explain next
(when $\alpha =0$,  the coupling functions are encoded in $F_{\Upsilon}(Y, \Upsilon)$ on the supergravity side, and hence still holomorphic
in supergravity variables).

Any affine special K\"ahler manifold $M$
can be realised as an immersion into a complex symplectic
vector space $V$ \cite{2002JGP....42...85A}, as we will review in section 
\ref{revaskm}. 
When passing from the prepotential $F^{(0)}$ to the generalized prepotential $F(Y, \Upsilon)$,
this construction gets extended, giving rise to a holomorphic family of immersions and deformed affine special K\"ahler manifolds, which combine
into a complex manifold $\hat{M} = M \times \mathbbm{C}$. 
By pulling back 
the standard Hermitian form of $V$, the space $\hat{M}$ becomes equipped
with a K\"ahler metric $g$ and a flat torsion free 
connection $\nabla$ which we use to define
special real coordinates. Taking the Legendre transform of the generalized
prepotential $F(Y,\Upsilon)$ as a Hesse potential, we can then define
a Hessian metric $g^H$. When analysing the integrability condition
for the existence of a Hesse potential, namely that $\nabla g^H$ must be
a completely symmetric rank three tensor in complex coordinates, one infers
the anomaly equation \eqref{anomal} with $\alpha =0$, as we will show in subsection \ref{hessianstr1}. 
This anomaly equation can also be viewed
as a consequence of a tension between preserving holomorphicity and symplectic covariance, as follows.
We introduce stringy variables ${\cal Y}^I$, and we derive various properties of the difference
$\Delta Y^I = {\cal Y}^I - Y^I$. We then express the symplectic function
$F_{\Upsilon} (Y, \Upsilon)$  in terms
of stringy variables. By taking multiple derivatives $\partial_{\Upsilon}|_{\cal Y}$
of $F_{\Upsilon}$ we obtain symplectic functions $\partial^n_{\Upsilon} F_{\Upsilon}|_{\cal Y} $ that we express back in terms of supergravity
variables. Then, setting $\Upsilon =0$, we obtain the set of symplectic 
functions $F^{(n)} = 2 i  \,  D_{\Upsilon}^{n-1} F_{\Upsilon}/n! 
|_{\Upsilon =0}$ 
introduced in \cite{deWit:1996ix} that, for $n \geq 2$, satisfy the anomaly equation \eqref{anomal} with $\alpha =0$. 
The symplectic covariant derivative $D_{\Upsilon}$ introduced in \cite{deWit:1996ix} is related to $\partial \Delta Y^I/\partial \Upsilon$, and thus it is simply a consequence of the passage
from stringy to supergravity variables. 
The
non-holomorphicity induced by the coordinate transformation reflects
the tension between holomorphicity and symplecticity, and is thus a universal feature
of the deformation induced by the passage from the prepotential $F^{(0)} (Y)$ to the generalized prepotential $F(Y, \Upsilon)$.

The aforementioned Hessian structure condition (namely that $\nabla g^H$ is totally symmetric)
gives  a master equation for
$F_\Upsilon$, 
\[
\partial_{\bar{I}} D_\Upsilon F_\Upsilon = \bar{F}^{JK}_I F_{\Upsilon J} F_{\Upsilon 
K} \;,
\]
which, upon applying $D^n_{\Upsilon}$ to it and setting $\Upsilon=0$, yields, by induction,  
 the anomaly
equation \eqref{anomal} with $\alpha =0$ for the functions  $F^{(n)}$ defined above.
This master equation for $F_\Upsilon$ is on par 
with the one derived for the topological 
free energy $F_{\rm top} $,
 \[
F_{\rm top}({\cal Y},\bar{\cal Y},Q) = \sum_{n=1}^{\infty} Q^n 
F^{(n)}({\cal Y}, \bar{\cal Y}) \;,
\]
where $Q$ is an expansion parameter related to the topological string
coupling, 
which satisfies 
\[
\partial_{\bar{I}} F_{\rm top} = i  \bar{F}^{(0)JK}_I \partial_J  F_{{\rm top}} \,
  \partial_K F_{{\rm top}} \;.
\]

Next, let us discuss the case $\alpha \neq 0$. When turning on $\alpha$,
$F_{\Upsilon}$ ceases to be holomorphic. Thus, starting from a non-holomorphic
generalized prepotential $F$ as in \cite{Cardoso:2014kwa}, we investigate the consequences for the master
equation for
$F_{\Upsilon}$ that result from the Hessian structure condition. The equation we obtain
is quite complicated.  To compare it with \eqref{anomal}, we specialise to a particular deformation,
proportional to $\alpha N^{IJ}_{(0)}$, where $N^{IJ}_{(0)}$ is the inverse 
of $N_{IJ}^{(0)} = - i (F^{(0)}_{IJ} - \bar{F}^{(0)}_{IJ})$. 
 Working at lowest order, we show that when setting $\Upsilon =0$, the master equation
for $F_{\Upsilon}$ equals the anomaly equation \eqref{anomal} with $n=2$. The anomaly equation for the higher $F^{(n)}$
can, in principle, be obtained from this master equation by acting with multiple covariant derivatives $D_{\Upsilon}$ on it.
Here,  $D_{\Upsilon}$ denotes the symplectic covariant derivative introduced in \cite{Cardoso:2012nh}, which is based on a non-holomorphic
generalized prepotential $F$.  We note that while the specific $\alpha$-deformation we  picked is tied to the topological string,
the framework presented in this paper is quite general and can be applied to other deformed systems, such as those discussed
in \cite{Cardoso:2012nh}.

The paper is organised as follows. In section \ref{skg} we review 
the extrinsic construction of special K\"ahler manifolds through immersion
into a model vector space.  In section \ref{hdspk} we
deform this construction by passing from the prepotential $F^{(0)}$ to the holomorphic
generalized prepotential $F(Y, \Upsilon)$. We introduce the Hessian structure based on special real coordinates, and use
the latter to introduce stringy variables $({\cal Y}^I, \Upsilon)$, as in \cite{Cardoso:2014kwa}. We relate the difference $\partial \Delta Y^I/\partial \Upsilon$ to the
symplectic covariant derivative of \cite{deWit:1996ix}, which we subsequently use
to derive a master equation for $F_{\Upsilon}$. Next, we use the Hessian structure to derive a different equation for
 $F_{\Upsilon}$, which we then relate to the holomorphic anomaly equation \eqref{anomal} with $\alpha =0$.
In section \ref{nholskg} we redo the analysis, but now based on a non-holomorphic generalized prepotential $F$.
In the concluding section we compare the approach of \cite{Cardoso:2014kwa} 
for obtaining the holomorphic anomaly equation with the approach taken here.
In the appendices we have collected some standard material to facilitate the reading of the paper.

\section{Review of special K\"ahler geometry \label{skg}}

\subsection{Affine special K\"ahler manifolds \label{revaskm} }

We start by reviewing the intrinsic definition of
(affine) special K\"ahler geometry given in 
\cite{Freed:1997dp}: A K\"ahler manifold $(M,g,\omega)$ 
with complex structure $J$ 
is affine special K\"ahler if there exists a flat, torsion-free,
symplectic connection $\nabla$ such that
\begin{equation}
\label{dNablaJ}
d_\nabla J=0 \;.
\end{equation}
We will refer to $\nabla$ as the special connection.
Our convention for the relation between metric $g$, K\"ahler form
$\omega$ and complex structure $J$ is 
\[
\omega ( \cdot \;, \cdot) = g ( \cdot \;, J \cdot ) \;,
\]
or, in local coordinates
\[
\omega_{ac} = g_{ab} J^b_{\;\;c} \;.
\]
The definition of the exterior covariant derivative
$d_\nabla$ is reviewed in appendix \ref{diff_geo}. 
As shown in appendix \ref{App:special_coord}, in $\nabla$-affine coordinates $q^a$ the condition 
(\ref{dNablaJ}) becomes
\begin{equation}
\label{curlJ}
\partial_{[a} J^b_{\;\;c]} = 0  \;,
\end{equation}
while the coefficients $\omega_{ab}$ are constant.
This in turn implies that
\[
\partial_ag_{bc} = \partial_b g_{ac}\;,
\]
which by applying the Poincar\'e lemma twice shows that the K\"ahler metric
is Hessian,
\[
g_{ab} = \partial^2_{a,b} H \;,
\]
where the real function $H$ is called a Hesse potential. The 
coordinate-free version of this local definition 
of a Hessian metric is as follows: given a Riemannian metric $g$ and
a flat, torsion-free connection $\nabla$, the pair $(g,\nabla)$ is
called a Hessian structure, and $g$ is called a Hessian metric, 
if the rank-3 tensor $\nabla g$ is totally
symmetric. 
It is easy to see that, given a 
K\"ahler manifold with a flat, torsion-free, symplectic connection
$\nabla$,  the condition
(\ref{dNablaJ}) is equivalent to the requirement that the metric 
$g$ is Hessian (that is $\nabla g$ is totally symmetric). 

On an affine special K\"ahler manifold 
one can choose the $\nabla$-affine coordinates $(q^a) = (x^I, y_I)$
to be Darboux coordinates, i.e. 
such that the K\"ahler form takes the standard form
\[
\omega = 2 dx^I \wedge dy_I = \Omega_{ab} dq^a \wedge dq^b \;,\;\;\;
(\Omega_{ab}) = \left( \begin{array}{cc}
0 & \mathbbm{1} \\
- \mathbbm{1} & 0 \\
\end{array} \right) \;.
\]
The coordinates $q^a$ are called special real coordinates,
and are unique up to affine transformations with symplectic
linear part.\footnote{We remark that the special coordinates $(q^a)$
differ from standard Darboux coordinates by a conventional normalization
factor, see appendix B for details.}

As shown in \cite{Freed:1997dp} the above definition is equivalent
to the well known alternative definition in terms of special holomorphic
coordinates $Y^I$ and of a holomorphic prepotential $F(Y^I)$.
We will now review the holomorphic formulation of special K\"ahler geometry
in the context of the universal extrinsic construction of 
\cite{2002JGP....42...85A}, which allows to realize any affine
special K\"ahler manifold, at least locally. For simply connected affine 
special K\"ahler manifold this construction in fact works globally. 
The universal construction realises special K\"ahler manifolds $M$
as immersions into the standard complex symplectic vector space
$V=T^* \mathbbm{C}^n$, where $\dim V = 2 \dim M = 4n$. We now review
some details, which we are going to use for our later generalised
construction.

Let $(Y^I, W_I)$ be complex Darboux coordinates
on $V=T^*\mathbbm{C}^n \simeq \mathbbm{C}^{2n}$. Then
\[
\Omega = dY^I \wedge dW_I
\]
is the standard complex symplectic form on $V$, and
\begin{equation}
\label{gammaV}
\gamma_V = i \Omega(\cdot, \overline{\cdot} ) = 
i \left( d Y^I \otimes d\overline{W}_I - d W_I \otimes
d \overline{Y}^I \right) = g_V + i \omega_V
\end{equation}
is the associated Hermitian form, $g_V$ is a flat (indefinite)
K\"ahler metric, and $\omega_V$ the corresponding K\"ahler form.

Next, let
\[
\phi: M \rightarrow V 
\]
be a non-degenerate, holomorphic, Lagrangian immersion 
of a complex manifold $M$
of (real) dimension $2n$ into $V$. We can assume, without loss
of generality, that the image $\phi(M)$ is realized as a graph, 
that is the immersion has been chosen such that, when identifying
$M$ locally with its image, we can take $Y^I$ as coordinates on the
locally embedded $M$, so that in terms of coordinates $(Y^I, W_I)$ 
the immersion takes the form 
\[
\phi: M \rightarrow V \;,\;\;\;(Y^I) \mapsto (Y^I, F_I(Y)) \;.
\]
This situation is generic, and can always be achieved, at least
locally, by a  symplectic transformation. 
Since the immersion $\phi$ is Lagrangian, we have
$\phi^*\Omega=0$, which is readily seen to be the integrability
condition for the existence of a holomorphic function $F$ such 
that $F_I = \partial F/\partial Y^I$. 
In the non-generic situation where $\phi(M)$ is not realized as
a graph, the immersion is still well defined and can be 
described using a complex symplectic vector $(Y^I(Z),W_I(Z))$,
where $Z=(Z^I)$ are holomorphic coordinates on $M$. However,
the components $Y^I$ cannot be used as coordinates on $M$, 
and the components $W_I$ fail to satisfy the integrability 
condition for the existence of a prepotential. This is 
well known in the literature as a symplectic vector (or
`holomorphic section') `without prepotential' \cite{Ceresole:1995jg}. We will 
assume in the following that we are in a generic symplectic
frame where a prepotential exists.

Since the immersion is non-degenerate, the pull back
$\gamma_M = \phi^* \gamma_V$ of the
Hermitian form $\gamma_V$ to $M$ is a non-degenerate 
Hermitian form, which by decomposition into real and
imaginary part defines a non-degenerate metric and two-form: 
\[
\gamma_M = 
g_M + i \omega_M \;.
\]
The explicit form of $\gamma_M$ is
\[
\gamma_M = N_{IJ} dY^I \otimes d\bar{Y}^J \;,
\]
where
\[
N_{IJ} = 2 \mbox{Im} F_{IJ} = - i (F_{IJ} - \bar{F}_{IJ}) = 
\frac{\partial^2}{\partial Y^I \partial \bar{Y}^J} [i (Y^K \bar{F}_K
- F_K \bar{Y}^K) ] \;.
\]
From this expression for $N_{IJ}$ it is manifest that
the metric $g_M$ is K\"ahler, with K\"ahler potential
\begin{equation}
\label{K_special}
K = i (Y^I \bar{F}_I - F_I \bar{Y}^I)  \;,
\end{equation}
and affine special K\"ahler with prepotential $F$.

Special holomorphic and special real coordinates are related 
as follows.
Given special holomorphic coordinates $Y^I$ on $M$, the corresponding
special real coordinate are given by the real part of the complex
symplectic vector $(Y^I, F_I)$:
\begin{eqnarray*}
Y^I &=& x^I + i u^I(x,y) \;, \\
F_I &=& y_I + i v_I(x,y) \;.
\end{eqnarray*}
Moreover, the holomorphic prepotential and the Hesse potential 
are related by a Legendre transform \cite{Cortes:2001}:
\[
H(x,y) = 2 \mbox{Im}F(x+iu(x,y)) - 2 y_I u^I(x,y)  \;.
\]
We remark that special real coordinates are well defined, at least
locally, in any symplectic frame (including those without a
prepotential) as a consequence of the non-degeneracy of the
symplectic form. For simply connected special K\"ahler manifolds
they are even globally defined functions, since the immersion 
is global, though not necessary global coordinates, since the immersion
need not be an embedding.

\subsection{Conical affine special K\"ahler manifolds}

While affine special K\"ahler manifolds are the scalar manifolds
of generic rigid $N=2$ vector multiplets, conical affine
special K\"ahler manifolds are the scalar manifolds of rigid
superconformal vector multiplets. These are in turn the
starting point for the construction of the coupling of 
vector multiplets to Poincar\'e supergravity, 
which proceeds as follows\footnote{This is reviewed 
in \cite{Mohaupt:2000mj,Freedman:2012zz}.}:
(i) start with a theory of $n+1$ superconformal vector multiplets,
(ii) gauge the superconformal algebra; this introduces various
connections which reside in the Weyl multiplet, (iii) partially gauge fix
the superconformal transformations to obain a theory of $n$ vector multiplets
coupled to Poincar\'e supergravity. In this construction the 
projective special K\"ahler manifold $\bar{M}$ of the supergravity theory
arises as a K\"ahler quotient of a conical affine special K\"ahler manifold.
Since we will not use this construction, we refer the interested reader
to the literature.

The additional condition implied by superconformal symmetry is, in 
terms of special holomorphic coordinates, that the prepotential
is homogeneous of degree two under complex scale transformations,
\[
F(\lambda X^I) = \lambda^2 F(X^I) \;,\;\;\; \lambda \in \mathbbm{C}^* \;.
\]
This is equivalent to the statement that the Hesse potential is 
homogeneous of degree two under real scale transformations of the
special real coordinates, and invariant under the $U(1)$ part of
$\mathbbm{C}^*$. The condition can also be formulated in a 
coordinate-free way \cite{2002JGP....42...85A}: 
a conical affine special K\"ahler manifold\footnote{Apart from 
`conical' the term `conic' is also use in the literature.}
is a special K\"ahler manifold equipped with a homothetic Killing
vector field $\xi$ satisfying
\[
\nabla \xi = D \xi  = \mbox{Id} \;,
\]
where $\nabla$ is the special connection, $D$ the Levi-Civita connection,
and Id the identity endomorphism on $TM$. One can then show that this
implies the existence of an infinitesimal holomorphic homothetic 
$\mathbbm{C}^*$ action on $M$, which is generated by $\xi$ and $J\xi$. 
To obtain a projective special K\"ahler
manifold by a K\"ahler quotient, one needs to assume that this group
action is free and proper.

\section{The holomorphic deformation \label{hdspk}}

\subsection{Deformation of the immersion}

One possible deformation of the vector multiplet action 
is to give it an explicit dependence on a background 
chiral multiplet \cite{deWit:1996ix}, see \cite{Mohaupt:2000mj} for
a review. By identifying this chiral multiplet
with the Weyl multiplet $W^2$, one can describe a particular 
class of higher derivative terms. Compatibility with 
superconformal symmetry determines the scaling behaviour
of the chiral multiplet, while insisting on a local supersymmetric
action implies that the dependence is holomorphic, that is
the standard F-term vector multiplet action is deformed by allowing the
prepotential, as a function on superspace, to depend explicitly on the chiral
multiplet. After integration over superspace, the action is a local
functional of the fields, which contains additional terms involving
holomorphic derivatives of the prepotential with respect to the background.
When identifying the chiral multiplet with the Weyl multiplet $W^2$,
one finds that the auxiliary fields cannot any longer be eliminated
in closed form, but only iteratively, thus generating an expansion
in derivatives. Such an action is naturally interpreted as a Wilsonian
effective action. 

In the following we will investigate how the introduction of a 
background field can be interpreted as a deformation of special
geometry. Since we focus on the scalar geometry, the background 
chiral field enters through its lowest component, a complex 
scalar denoted $\Upsilon$. The generalized prepotential 
$F(Y,\Upsilon)$ is holomorphic in $Y^I$ and $\Upsilon$, and
(graded) homogeneous of degree two, that is
\[
F(\lambda Y, \lambda^w \Upsilon) = \lambda^2 F(Y,\Upsilon) \;,\;\;\;
\lambda \in \mathbbm{C} \;,
\]
where $w$ is the weight of $\Upsilon$ under scale transformations.
If $\Upsilon$ is the lowest component of the Weyl multiplet $W^2$, then
$w=2$. Our geometric model for the deformation parameterized by 
$\Upsilon$ is a map
\begin{equation}
\label{phi_holo}
\phi: \hat{M} := M \times \mathbbm{C} \rightarrow V \;,\;\;\;
(Y^I, \Upsilon) \mapsto (Y^I, F_I (Y, \Upsilon)) \;,
\end{equation}
which can be interpreted as a holomorphic family of immersions
$\phi_\Upsilon :  M \rightarrow V \;,\;\; (Y^I) \mapsto (Y_I,F_I(Y,\Upsilon))$,
that define a family of affine special K\"ahler structures on $M$. 
While
$\Upsilon$ is a scalar under symplectic transformations, it enters
into the transformation of the 
complex symplectic vector $(Y^I, F_I(Y,\Upsilon))$, and other 
objects, through the generalized prepotential. 
Our set-up is consistent
with \cite{deWit:1996ix}, in particular we can draw on the various formulae
for symplectic transformations derived there.

We define a metric and a two-form on $\hat{M}=M\times\mathbbm{C}$
by pulling back the canonical hermitian form $\gamma_V$:
\[
\gamma = \phi^* \gamma_V = g + i \omega = 
N_{IJ} dY^I \otimes d\bar{Y}^J + i \bar{F}_{I\Upsilon}
dY^I \otimes d\bar{\Upsilon} - i F_{I\Upsilon} d\Upsilon \otimes
d\bar{Y}^I \;,
\]
where $N_{IJ}= -i (F_{IJ} - {\bar F}_{IJ})$. 
We assume that $\gamma$ is non-degenerate, which certainly 
is true for sufficiently small $\Upsilon$.\footnote{In applications
$\Upsilon$ will not necessarily be small, but it is reasonable to 
expect that $\gamma$ is non-degenerate, at least generically.}
In the following, holomorphic coordinates on $\hat{M}$ are
denoted $(v^A) = (Y^I,\Upsilon)$. Using the conventions
\begin{eqnarray}
da \, db &=& \frac12 \left( da \otimes db + db \otimes da \right) \;, \nonumber\\
da \wedge db &=& da \otimes db - db \otimes da \:, \nonumber
\end{eqnarray}
we obtain the 
metric 
\[
g = g_{AB} d v^A d\bar{v}^{{B}} =
N_{IJ} dY^I d\bar{Y}^J 
+ i \bar{F}_{I\Upsilon} dY^I d\bar{\Upsilon}
- i F_{J \Upsilon} d \Upsilon d \bar{Y}^J \;,
\]
which is a K\"ahler metric  $g_{AB} = \partial^2_{A,\bar{B}} K$
with 
K\"ahler potential 
\begin{equation}
K = - i \left( \bar{Y}^I F_I(Y,\Upsilon) - \bar{F}_I(\bar{Y}, \bar{\Upsilon}) 
Y^I \right) \;,
\label{K_ups}
\end{equation}
and
\[
\omega = -\frac{i}{2} N_{IJ} dY^I \wedge d\bar{Y}^J + \frac{1}{2}
\bar{F}_{I\Upsilon} dY^I \wedge d\bar{\Upsilon} - \frac{1}{2}
F_{I\Upsilon} d\Upsilon \wedge d \bar{Y}^I
\]
is the associated K\"ahler form. The K\"ahler metric $g_{AB}$
has occured in the deformed sigma model \cite{Cardoso:2012mc},
which provides a field theoretic realization of our set-up. 

\subsection{Real coordinates and the Hesse potential}

Following \cite{LopesCardoso:2006bg}, we now define special real
coordinates and a Hesse potential in presence of the deformation.
Special real coordinates are defined by
\[
Y^I = x^I + i u^I(x,y,\Upsilon, \bar{\Upsilon}) \;,\;\;\;
F_I = y_I + i v_I(x,y, \Upsilon, \bar{\Upsilon}) \;,
\]
and the (generalized) Hesse potential is related to the 
(generalized) prepotential by a Legendre transform:
\[
H(x,y,\Upsilon, \bar{\Upsilon}) =
- i (F-\bar{F}) - 2 y_I u^I(x,y,\Upsilon,\bar{\Upsilon}) \;,
\]
where $F=F(Y(x,u(x,y,\Upsilon,\bar{\Upsilon})), \Upsilon)$. 

We are interested in the coordinate transformation between special complex and special real coordinates\footnote{We find
it convenient to work with $\Upsilon$ and $\bar{\Upsilon}$ when 
using special `real' coordinates instead of decomposing them 
into their real and imaginary parts.} 
\[
(x,u,\Upsilon, \bar{\Upsilon}) \mapsto
(x,y(x,u,\Upsilon,\bar{\Upsilon}), \Upsilon, \bar{\Upsilon})
\]
and its inverse
\[
(x,y,\Upsilon,\bar{\Upsilon}) \mapsto 
(x,u(x,y,\Upsilon, \bar{\Upsilon}), \Upsilon, \bar{\Upsilon}) \;.
\]

When rewriting derivatives between the coordinate systems,
one needs to carefully use the chain rule: when differentiating
a function $f=f(x,y(x,u,\Upsilon,\bar{\Upsilon}), \Upsilon, \bar{\Upsilon})$
the following formulae are useful
\begin{eqnarray*}
\left. \frac{\partial f}{\partial x^I} \right|_{u} &=& 
\left. \frac{\partial f}{\partial x^I} \right|_{y} +
\left. \frac{\partial f}{\partial y_K} \right|_{x} 
\frac{\partial y_K}{\partial x^I}\;,  \\
\left. \frac{\partial f}{\partial u^I} \right|_{x} &=&
\left. \frac{\partial f}{\partial y_K}  \right|_{x} 
\frac{\partial y_K}{\partial u^I} \;,  \\
\left. \frac{\partial f}{\partial \Upsilon} \right|_{x,u} &=&
\left. \frac{\partial f}{\partial \Upsilon} \right|_{x,y} +
\left. \frac{\partial f}{\partial y_K} \right|_{x} 
\frac{\partial y_K}{\partial \Upsilon} \;.
\end{eqnarray*}
The Jacobians for the coordinate transformations take the form
\[
\frac{D(x,u,\Upsilon, \bar{\Upsilon})}{D(x,y,\Upsilon,\bar{\Upsilon})} =
\left( \begin{array}{cccc}
\mathbbm{1} & 0 & 0 & 0 \\
\left. \frac{\partial u}{\partial x} \right|_y &
\left. \frac{\partial u}{\partial y} \right|_x &
\left. \frac{\partial u}{\partial \Upsilon} \right|_{x,y} &
\left. \frac{\partial u}{\partial \bar{\Upsilon}} \right|_{x,y} \\
0 & 0 & \mathbbm{1} & 0 \\
0 & 0 & 0 & \mathbbm{1} \\
\end{array} \right)
\]
and
\[
\frac{D(x,y,\Upsilon, \bar{\Upsilon})}{D(x,u,\Upsilon,\bar{\Upsilon})} =
\left( \begin{array}{cccc}
\mathbbm{1} & 0 & 0 & 0 \\
\left. \frac{\partial y}{\partial x} \right|_u &
\left. \frac{\partial y}{\partial u} \right|_x &
\left. \frac{\partial y}{\partial \Upsilon} \right|_{x,u} &
\left. \frac{\partial y}{\partial \bar{\Upsilon}} \right|_{x,u} \\
0 & 0 & \mathbbm{1} & 0 \\
0 & 0 & 0 & \mathbbm{1} \\
\end{array} \right) \;.
\]
By the chain rule it is straightforward to evaluate
\[
\frac{D(x,y,\Upsilon,\bar{\Upsilon})}{D(x,u,\Upsilon,\bar{\Upsilon})} =
\left( \begin{array}{cccc}
\mathbbm{1}  & 0 & 0 & 0 \\
\frac{1}{2} R & - \frac{1}{2} N & \frac{1}{2} F_{I\Upsilon} &
\frac{1}{2} \bar{F}_{I\Upsilon} \\
0 & 0 & \mathbbm{1} & 0 \\
0 & 0 & 0 & \mathbbm{1} \\
\end{array} \right) \;,
\]
where $2 F_{IJ} = R_{IJ} + i N_{IJ}$. 
This matrix can easily be inverted, with the result:
\[
\frac{D(x,u,\Upsilon,\bar{\Upsilon})}{D(x,y,\Upsilon,\bar{\Upsilon})} =
\left( \begin{array}{cccc}
\mathbbm{1} & 0 & 0 & 0 \\
N^{-1} R & -2 N^{-1} & N^{-1} F_{I\Upsilon} & N^{-1} \bar{F}_{I\Upsilon} \\
0 & 0 & \mathbbm{1} & 0 \\
0 & 0 & 0 & \mathbbm{1} \\
\end{array} \right) \;.
\]
In order to transform the K\"ahler metric to special real
coordinates, the following relations are useful:
\[
\frac{\partial H}{\partial x^I} = 2 v_I \;,\;\;\;
\frac{\partial H}{\partial y_I} = - 2 u^I \;.
\]
Moreover, using the chain rule one computes: 
\begin{eqnarray}
\left. \frac{\partial v_I}{\partial x^J}\right|_y  &=&
\frac{1}{2} \left( N + RN^{-1}R \right)_{IJ} \;, \nonumber\\
\left. \frac{\partial v_I}{\partial y_J}\right|_x  &=& \left. - \frac{\partial u^J}{\partial x^I}\right|_y = 
2 \left( N^{-1} \right)^{IJ} \;, \nonumber\\
\left. \frac{\partial v_I}{\partial u^J}\right|_x &=& \frac{1}{2} R_{IJ} \;. \nonumber
\end{eqnarray}
Using the notation $(q^a)=(x^I, y_I)$, the K\"ahler metric $g$
expressed in special real variables takes the 
form
\[
g = \frac{\partial^2 H}{\partial q^a \partial q^b} dq^a dq^b
+ \frac{\partial^2 H}{\partial q^a \partial \Upsilon} dq^a d\Upsilon
+ \frac{\partial^2 H}{\partial q^a \partial \bar{\Upsilon}} dq^a 
d\bar{\Upsilon} \;,
\]
where 
\[
\left( \frac{\partial^2 H}{\partial q^a \partial q^b} \right)
= \left( \begin{array}{cc}
N + RN^{-1} R & -2 RN^{-1} \\
- 2 N^{-1} R & 4 N^{-1} \\
\end{array} \right) \;,
\]
and 
\[
\frac{\partial^2 H}{\partial x^I \partial \Upsilon} =
2 \bar{F}_{IM} N^{MN} F_{N\Upsilon} \;,\;\;\;
\frac{\partial^2 H}{\partial x^I \partial \bar{\Upsilon}} =
2 {F}_{IM} N^{MN} \bar{F}_{N\Upsilon} \;,\;\;\;
\]
\[
\frac{\partial^2 H}{\partial y_I \partial \Upsilon} =
-2 N^{IJ} F_{J\Upsilon} \;,\;\;\;
\frac{\partial^2 H}{\partial y_I \partial \bar{\Upsilon}} =
-2 N^{IJ} \bar{F}_{J\Upsilon} \;.
\]
In the undeformed case the K\"ahler metric is simultaneously Hessian.
To see whether this is still the case, we first note that $\hat{M}$
can be equipped with an affine structure and thus a Hessian metric
$g^H$ with Hesse potential $H$ can be defined. 
This  requires the existence of a flat, torsion-free
connection. For fixed $\Upsilon$ we know that the special 
connection $\nabla$ is such a connection, with affine coordinates 
$x^I,y_I$. We can extend $\nabla$ to a
flat, torsion-free connection on $\hat{M}=M\times \mathbbm{C}$ by
imposing 
\[
\nabla dx^I = 0 \;,\;\;\;
\nabla dy_I= 0 \;,\;\;\;
\nabla d\Upsilon = 0 \;,\;\;\;
\nabla d\bar{\Upsilon} = 0 \;.
\]
If $x^I, y_I$ are not global coordinates on $M$, we use that
$M$ can be covered by special real coordinate systems, which
are related by affine transformations with symplectic linear part.
Since for fixed  $\Upsilon\not=0$ the map $\phi_\Upsilon$ still induces an
affine special K\"ahler structure, special real coordinate systems
extend to $\hat{M}$ and provide it with the affine structure required
to define a flat torsion-free connection.

Upon computing the components of the Hessian metric $g^H$ explicitly,
we realize that is not equal to the K\"ahler metric $g$. 
The difference between the two metrics is
\[
g^H - g =\partial^2 H_{|x,y} = 
\frac{\partial^2 H}{\partial \Upsilon \partial \Upsilon} d\Upsilon d\Upsilon
+ 2 \frac{\partial^2 H}{\partial \Upsilon \partial \bar{\Upsilon}} 
d\Upsilon d\bar{\Upsilon} 
+ \frac{\partial^2 H}{\partial \bar{\Upsilon} \partial \bar{\Upsilon}} 
d\bar{\Upsilon} d\bar{\Upsilon} \;,
\]
where
\[
\frac{\partial^2 H}{\partial \Upsilon \partial \bar{\Upsilon}} =
N^{IJ} F_{I\Upsilon} \bar{F}_{J\Upsilon} \;,\;\;\;
\frac{\partial^2 H}{\partial \Upsilon \partial \Upsilon} =
-i F_{\Upsilon \Upsilon} + N^{IJ} F_{I\Upsilon} F_{J\Upsilon} \;,\;\;\;
\]
\[
\frac{\partial^2 H}{\partial \bar{\Upsilon} \partial \bar{\Upsilon}} =
i \bar{F}_{\Upsilon \Upsilon} + N^{IJ} \bar{F}_{I\Upsilon} \bar{F}_{J\Upsilon} 
\;.
\]
We remark that these metric coefficients are symplectic 
functions, see \cite{deWit:1996ix}, which is necessary 
in order that $g^H - g$ is a well defined tensor field
(which we know to be the case, because $g^H$ and $g$ are 
both metric tensors). We further remark that
\[
2 H = K -2i  \Upsilon F_\Upsilon + 2i \bar{\Upsilon} \bar{F}_\Upsilon
\]
differs from the K\"ahler potential (\ref{K_ups}) 
by a K\"ahler transformation. Therefore $2H$, taken as a 
K\"ahler potential, defines the same K\"ahler metric 
$g=g^K$ 
as $K$. However, when taking $K$ as a Hesse potential
one does not get the Hessian metric $g^H$.
While a K\"ahler potential is
unique up to K\"ahler transformations, a Hesse potential
is unique up to affine transformations. Moreover, since
our Hessian metric has definite scaling properties, we can
impose that the Hesse potential is homogeneous of
degree two, which is automatic in the way we have defined it
as the Legendre transform of the generalized prepotential.
If homogeneity is imposed on 
top of using special real coordinates, then the
Hesse potential is unique up to symplectic transformations. 
We remark that the Hesse potential is the sum of 
two symplectic functions. Different linear combinations of
these two functions define different metrics. By inspection
one finds that defining the (generalized) Hesse potential as the Legendre 
transform of the generalized prepotential 
leads to a particularly simple form 
of the coefficients $\partial^2H_{|x,y}$. We will see later 
how the Hessian metric $g^H$ encodes the holomorphic
anomaly equation. 

\subsection{Deformed special K\"ahler geometry}

We are now in position to demonstrate that $\hat{M}$ 
carries itself a deformed version of affine special K\"ahler geometry.
We have already seen that $g$ is a K\"ahler metric with 
K\"ahler form $\omega$. To compare this with the two-form 
$2 dx^I \wedge dy_I$, which is the K\"ahler form on $M$,
we compute
\begin{eqnarray}
2 dx^I \wedge dy_I &=&
- \frac{i}{2} N_{IJ} dY^I \wedge d\bar{Y}^J -
\frac{1}{2} F_{I\Upsilon} d \Upsilon \wedge d\bar{Y}^I +
\frac{1}{2} \bar{F}_{I \Upsilon} dY^I \wedge d\bar {\Upsilon} 
\nonumber\\
&& +
\frac{1}{2} F_{I \Upsilon} dY^I \wedge d\Upsilon +
\frac{1}{2} \bar{F}_{I\Upsilon} d\bar{Y}^I \wedge d \bar{\Upsilon} \;,
\label{xyold}
\end{eqnarray}
and therefore the K\"ahler form can be written as
\[
\omega = 2 dx^I \wedge dy_I 
- \frac{1}{2} F_{I\Upsilon} dY^I \wedge d\Upsilon
- \frac{1}{2} \bar{F}_{I \Upsilon} d\bar{Y}^I \wedge d\bar{\Upsilon} \;.
\]
This shows in particular that $2 dx^I \wedge dy_I$, when considered
as a form on $\hat{M}$,
is not of type $(1,1)$ (since $\omega$ is, and both differ by pure forms). Using
the rewriting
\[
F_{I \Upsilon} dY^I \wedge d \Upsilon =
d F_\Upsilon \wedge d \Upsilon = 
- d (\Upsilon d F_{\Upsilon}) \;,
\]
we find
\begin{equation}
\label{Diff_holo}
\omega = 2 dx^I \wedge dy_I  + \frac{1}{2} d(\Upsilon d F_\Upsilon) +
\frac{1}{2} d ( \bar{\Upsilon} d\bar{F}_{\Upsilon}) \;.
\end{equation}
Thus the difference between the K\"ahler forms
$\omega$ of $\hat{M}$ and $2dx^I \wedge dy_I$ of $M$ is 
exact, so that both forms are homologous. The deformation 
involves the function $F_\Upsilon=\partial_\Upsilon F$, which plays
a central role in describing the deformation and should be viewed
as the supergravity 
counterpart of the topological free energy $F_{\rm top}$.
First, note that while the generalized prepotential $F$, and 
its higher derivatives $\partial^n_{\Upsilon} F$ with $n>1$, are not
symplectic functions, $F_\Upsilon$ is a symplectic function \cite{deWit:1996ix}. 
Moreover, it is independent of the 
undeformed (two-derivative) prepotential $F^{(0)}(Y) = F(Y,\Upsilon=0)$,
but contains all the information about the deformation. We remark that
while within the present construction $F_\Upsilon$ is holomorphic, this
condition will be relaxed later.

Next we compute
\begin{equation}
\label{DeformOmega}
\nabla \omega = - \frac{1}{2} d (F_{I\Upsilon}) \otimes (dY^I \wedge
d \Upsilon) + c.c.
\end{equation}
which shows that $\omega$ is not parallel, and the connection $\nabla$
is not a symplectic connection on $\hat{M}$. This shows that while
$(\hat{M}, g, \omega, \nabla)$ is K\"ahler, it is not special 
K\"ahler. The deformation is controlled by an exact form, which 
is determined by the symplectic function $F_\Upsilon$.

The fourth condition on a special connection is that the complex
structure is covariantly closed. 
To compute the exterior covariant derivative of the complex structure
$J$, we note that the vector fields $\partial_{x^I}, \partial_{y_I},
\partial_\Upsilon , \partial_{\bar{\Upsilon}}$ define a $\nabla$-parallel
frame which is dual to the $\nabla$-parallel co-frame 
$dx^I, dy_I, d\Upsilon, d\bar{\Upsilon}$. Using this one verifies that
\[
\nabla \frac{\partial}{\partial Y^I} = 
\nabla \left( \frac{1}{2} \frac{\partial}{\partial x^I} +
\frac{1}{2} F_{IJ} \frac{\partial}{\partial y_J} \right) =
\frac{1}{2} d F_{IJ} \otimes \frac{\partial}{\partial y_J} \;.
\]
Using that $d_\nabla J = dJ^a e_a - J^a \wedge d_\nabla e_a$
where $e_a$ is any basis of sections of $T\hat{M}$, 
so that $d_\nabla e_a = \nabla e_a$, 
we find
\[
d_\nabla J = \left( - i d Y^I \wedge \frac{1}{2} d F_{IJ} 
+ c.c. \right) 
\otimes \frac{\partial}{\partial y_J}  \;.
\]
Note the rewriting
\[
dY^I \wedge dF_{IJ} = dY^I \wedge F_{IJ\Upsilon} d\Upsilon = 
- d(F_{IJ} d Y^I ) = d(F_{I\Upsilon} d\Upsilon) \;,
\]
where we used symmetry of $F_{IJ}$ and the chain rule. 
Therefore
\begin{equation}
\label{DeformJ}
d_\nabla J = \left( -i d(F_{I\Upsilon} d\Upsilon) + c.c.\right)
\otimes \frac{\partial}{\partial y_I} =
\left( - i F_{IJ\Upsilon} dY^J \wedge d\Upsilon + c.c. \right)
\otimes \frac{\partial}{\partial y_I} \;,
\end{equation}
which is non-vanishing.
As a consistency check, observe that it is manifest that
$d_\nabla^2=0$, which must be true because $\nabla$ is flat.
Since the complex structure $J$ of $\hat{M}$ is not 
covariantly closed, the fourth condition required on the
connection $\nabla$ in order to define a special K\"ahler
manifold is also violated. Again the deformation involves
an exact form constructed out of the function $F_\Upsilon$.

In summary, $(\hat{M} = M \times \mathbbm{C}, J, g)$ is a K\"ahler 
manifold with K\"ahler form $\omega$, 
equipped with a flat, torsion-free connection, such that $\nabla \omega$
and $d_\nabla J$ are given by (\ref{DeformOmega}) and (\ref{DeformJ}). 
We will call such manifolds deformed affine special K\"ahler manifolds.
Since our definition involves the map $\phi$, this is not an intrinsic
definition, but the name for a specific construction.

For completeness we remark that the pullback of the complex 
symplectic form $\Omega$ of $V$ is non-vanishing:\footnote{It is of
course clear already for dimensional reasons that
$\hat{M}$ cannot be a (locally immersed) Lagrangian submanifold
of $V$.}
\[
\phi^* \Omega = F_{I\Upsilon} dY^I \wedge d\Upsilon =
- d(\Upsilon d F_\Upsilon) \;.
\]
As we by now expect, the right hand side is exact and controlled by $F_\Upsilon$. 


\subsection{Stringy complex coordinates \label{scc}}

The framework introduced so far is based on a generalized
holomorphic prepotential $F(Y,\Upsilon)$, a complex symplectic vector
$(Y^I, F_I(Y,\Upsilon))$ and a map $\phi: \hat{M} \rightarrow V$
which introduces a K\"ahler metric $g$ on $\hat{M} = M \times \mathbbm{C}$,
which deviates from being special K\"ahler if $F_\Upsilon \not=0$. 
Although $\Upsilon$ is a symplectic scalar, symplectic transformations
of data derived from $F$ or the symplectic vector $(Y^I, F_I)$
depend on $\Upsilon$.
If we expand $F$ in a power series
\[
F(Y,\Upsilon) = \sum_{g=0}^\infty F^{(g)}(Y) \Upsilon^g \;,
\]
then the functions $F^{(g)}(Y)$  are holomorphic and 
homogeneous of degree $2-2g$, but they are not symplectic 
functions, and transform in a complicated way under symplectic
transformations.

When the background is identified with the Weyl multiplet $W^2$, our
formalism describes an Wilsonian effective action for vector
multiplets which includes a certain class of higher derivative
terms.  The same class of terms can be described using the topological 
string, but the formalism used in this context is different. There is no 
generalized prepotential, but instead one works with an undeformed 
complex symplectic vector $({\cal Y}^I, F^{(0)}_I({\cal Y}))$.
The information which is encoded in the symplectic function 
$F_\Upsilon$ in the supergravity formalism is then differently 
encoded in a hierarchy 
of genus-$g$ topological free energies
$F^{(g)}({\cal Y}, \bar{\cal Y})$ which individually are symplectic functions,
at the expense of not being holomorphic. The deviation from
holomorphicity is controlled by the holomorphic anomaly 
equation. Elaborating on \cite{Cardoso:2014kwa}, we will now show
that the relation between the two frameworks can be understood as
a coordinate transformation. This will proceed in two steps. First
we will show that when starting from the holomorphically deformed
special geometry introduced so far, one obtains a hierarchy of
free energies, where $F^{(1)}$ is holomorphic, while the $F^{(g)}$
with $g>1$ are non-holomorphic and satisfy a version of the
holomorphic anomaly equation where the two-derivative term
is absent. This is not quite the situation for the topological string,
where already $F^{(1)}$ is non-holomorpic and 
the anomaly equation requires an additional two-derivative term.
In the next section we will generalize our deformed
special geometry by making it explicitly non-holomorphic, and then
show that by a coordinate transformation we obtain 
the full anomaly equation.

The relation between the supergravity coordinates $Y^I$ and 
the stringy coordinates ${\cal Y}^I$ is defined 
by imposing that the corresponding special real coordinates
agree \cite{Cardoso:2014kwa}:
\[
\left( \begin{array}{c}
2 x^I \\
2 y_I \\
\end{array} \right) =
\left( \begin{array}{c}
Y^I + \bar{Y}^I \\
F_I(Y,\Upsilon) + \bar{F}_I(\bar Y, \bar \Upsilon) \\
\end{array} \right) = 
\left( \begin{array}{c}
{\cal Y}^I + \bar{\cal Y}^I \\
F^{(0)}_I ({\cal Y}) + \bar{F}^{(0)}_I(\bar {\cal Y}) \\
\end{array} \right) \;.
\]
This implicitly defines a non-holomorphic coordinate 
transformation between complex coordinates on $\hat{M}$,
\begin{equation}
(Y^I, \Upsilon) \mapsto ({\cal Y}^I, \Upsilon) \;,
\label{oldnewc}
\end{equation}
which we parametrize as \cite{Cardoso:2014kwa}
\[
{\cal Y}^I = Y^I + \Delta Y^I(Y,\bar{Y},\Upsilon, \bar{\Upsilon}) \;.
\]
Note that by construction ${\cal Y}^I = Y^I$ for $\Upsilon=0$.
In particular, the ${\cal Y}^I$ still are holomorphic coordinates on $M$. 
If $\Upsilon \not=0$ the coordinate transformation can be 
constructed iteratively \cite{Cardoso:2014kwa}. Since this gets
complicated very soon, we will focus on statements that can be
made without expansion or iteration.

To this end, let us consider the two-form $2dx^I \wedge dy_I$ given in 
\eqref{xyold}. Using \eqref{oldnewc}, we express this two-form in the new complex variables $({\cal Y}^I,
\bar{\cal Y}^I)$, obtaining
\begin{eqnarray}
2 dx^I \wedge dy_I &=&
- \frac{i}{2} N_{IJ} \left[ - \frac{\partial \Delta Y^J}{\partial {\cal Y}^K} \, d {\cal Y}^K \wedge d {\cal Y}^I
- \frac{\partial \Delta Y^J}{\partial {\bar{\cal Y}}^K} \, d {\bar{\cal Y}^K} \wedge d {\bar {\cal Y}}^I \right.
\nonumber\\
&& \qquad \qquad \left. + \left( \delta^I_K \delta^J_L -  \frac{\partial \Delta Y^I}{\partial {{\cal Y}}^K} \delta^J_L + \delta^I_K 
\frac{\partial \Delta Y^J}{\partial {\bar {\cal Y}}^L} \right)
d{\cal Y}^K \wedge d\bar{\cal Y}^L \right. \nonumber\\
&& \qquad \qquad \left.  + 
2 \frac{\partial \Delta Y^I}{\partial \Upsilon} \, dx^J \wedge d \Upsilon
+ 2 
\frac{\partial \Delta Y^I}{\partial \bar \Upsilon} \, dx^J \wedge d {\bar \Upsilon}
\right]
\nonumber\\
&& +
F_{I \Upsilon} dx^I \wedge d\Upsilon +
\bar{F}_{I\Upsilon} dx^I \wedge d \bar{\Upsilon} \;,
\label{xynew}
\end{eqnarray}
where in the last two lines we combined various terms into terms containing $d x^I$.
We now convert all differentials appearing in \eqref{xynew} to the real flat frame
$(d x^I, dy_I, d \Upsilon, d \bar \Upsilon)$ using
\[
d {\cal Y}^I = dx^I + i N_{(0)}^{IK} R^{(0)}_{KJ} \, d x^J - 2 i N^{IJ}_{(0)} \, d y_J \:,
\]
where here and in the following we use a notation where the subscript or
superscript $(0)$ indicates that a quantity has been calculated 
using the undeformed prepotential $F^{(0)}({\cal Y}) = F(Y,\Upsilon=0)$. 
Then, by comparing the differentials on both sides of the resulting expression, 
we obtain the relations 
\begin{eqnarray}
\frac{\partial \Delta Y^J}{\partial \Upsilon} &=& 
- i N^{JK} F_{K\Upsilon} \label{R1} \;, \\
N^{IJ} &=& \left( \delta^I_K - \frac{\partial \Delta Y^I }{\partial {\cal 
Y}^K}  + \frac{\partial \Delta Y^I}{\partial \bar{\cal Y}^K} \right)
N^{KJ}_{(0)} \;, \nonumber\\
N_{[IK} \frac{\partial \Delta Y^K}{\partial {\cal Y}^{J]}} & = & 0 \;, \nonumber
\label{relDelY}
\end{eqnarray}
where in the last equation the square bracket denotes antisymmetrization of the uncontracted
indices.

\subsection{The symplectic covariant derivative}

The advantage of the stringy coordinates is that the variable
$\Upsilon$ does not enter into symplectic transformations. Thus
given any symplectic function $G({\cal Y},\Upsilon)$ (not necessarily 
holomorphic), then the symplectic transformation behaviour
is not modified when taking partial derivatives with respect to 
$\Upsilon$. In particular, if $G({\cal Y}, \Upsilon)$ is a symplectic 
function, then so is $\partial G/\partial \Upsilon$. In contrast,
when using the supergravity variables $Y^I$, then partial derivatives
with respect to $\Upsilon$ change the symplectic transformation 
behaviour. For example, while the derivative $F_\Upsilon$ of the
generalized prepotential is a symplectic function, its derivatives
like $F_{\Upsilon \Upsilon}$ are not \cite{deWit:1996ix}. Using (\ref{R1}) there is a systematic
way to compensate for this behaviour. Suppose $G(Y,\Upsilon)$ is 
a symplectic function, given in supergravity variables, which for
the time being we assume to be holomorphic.\footnote{This restriction
will be lifted later.} Expressing $G$ in stringy variables, we 
obtain $G=G(Y({\cal Y}, \Upsilon), \Upsilon)$.\footnote{Note that,
though we do not indicate this by notation, 
$Y({\cal Y},\Upsilon)$ is not holomorphic.} 
When regarding $G$ as a function of ${\cal Y}$ and $\Upsilon$, the
partial derivative with respect to $\Upsilon$ 
is again a symplectic function. Now apply 
the chain rule:
\[
\left. \frac{\partial G}{\partial \Upsilon}\right|_{{\cal Y}} =
\left. \frac{\partial G}{\partial \Upsilon}\right|_{Y} +
\left. \frac{\partial G}{\partial Y^I} \right|_{\Upsilon}
\frac{\partial Y^I}{\partial \Upsilon}
\]
and use (\ref{R1})
\[
\frac{\partial Y^I}{\partial \Upsilon} = - \frac{\partial \Delta Y^I}{
\partial \Upsilon} = + i N^{IJ} F_{J\Upsilon} \;.
\]
Therefore, if $G(Y,\Upsilon)$ is a symplectic function, then 
\[
D_\Upsilon G  := 
\left. \frac{\partial G}{\partial \Upsilon}\right|_{Y} 
+ i N^{IJ} F_{J\Upsilon} \frac{\partial G}{\partial Y^I} 
\]
is also symplectic.
The expression 
\begin{equation}
\label{D_Upsilon_holo}
D_\Upsilon = \left. \frac{\partial }{\partial \Upsilon}\right|_{Y} 
+ i N^{IJ} F_{J\Upsilon} \frac{\partial }{\partial Y^I} \;,
\end{equation}
which we have derived from the coordinate transformation between
supergravity and stringy variables, is the symplectic covariant
derivative which was introduced in \cite{deWit:1996ix} based on
studying the symplectic transformation behaviour of derivatives
of symplectic functions. We remark that while $G$ was assumed
holomorphic, $D_\Upsilon G$ is not holomorphic, due to the presence
of the inverse metric $N^{IJ}$. By taking higher covariant derivatives
$D^n_\Upsilon G$, one can create a whole tower of symplectic functions.
We remark that when the initial function $G$ is non-holomorphic, the
covariant derivative needs to be modified, as will be discussed later.

The main application of this result is to show how one can obtain,
starting from $F_\Upsilon(Y,\Upsilon)$ a hierarchy of functions
$F^{(n)}({\cal Y}, \bar{\cal Y})$ which can be interpreted as 
topological free energies, because they satisfy the holomorphic
anomaly equation. While this is result known from \cite{deWit:1996ix}
we briefly explain how this works and how the hierarchy of
equations for the functions $F^{(n)}({\cal Y}, \bar{\cal Y})$  
can be consolidated into a master anomaly 
equation.

First, following \cite{deWit:1996ix} we define\footnote{Note that
here we use a different normalisation from the one used in section \ref{intro}.} a hierarchy of 
symplectic functions through covariant derivatives of the holomorphic
symplectic function $F_\Upsilon(Y,\Upsilon)$: 
\[
\Phi^{(n)}(Y,\bar{Y},\Upsilon, \bar{\Upsilon}) 
= \frac{1}{n!} D^{n-1}_\Upsilon F_\Upsilon \;,
\]
for $n=1,2,\ldots$, and $\Phi^{(0)}=0$. Then
\begin{eqnarray*}
\Phi^{(1)} & = & F_\Upsilon \\
\Phi^{(2)} & = & \frac{1}{2} D_\Upsilon F_\Upsilon \;,
\end{eqnarray*}
etc. $\Phi^{(1)}$ is the only holomorphic function in this hierarchy. 
One computes 
\[
\frac{\partial \Phi^{(2)}}{\partial \bar{Y}^I} = 
\frac{i}{2} \frac{\partial N^{JK}}{\partial \bar{Y}^I} F_{J\Upsilon} 
F_{K\Upsilon} = \frac{1}{2} \bar{F}_{I}^{JK} \partial_J \Phi^{(1)} \partial_K
\Phi^{(1)} \;,
\]
where $ \bar{F}_{I}^{JK} =  \bar{F}_{IPQ} N^{PJ} N^{QK}$.
From this starting point it is straightforward to obtain 
the holomorphic anomaly equation 
\begin{equation}
\label{A0}
\frac{\partial \Phi^{(n)}}{\partial \bar{Y}^I} = \frac{1}{2} 
\bar{F}_{I}^{JK} \sum_{r=1}^{n-1} \partial_J \Phi^{(r)} \partial_K
\Phi^{(n-r)} \;,\;\;n\geq 2
\end{equation}
by complete induction. Next we define
\begin{equation}
F^{(n)}({\cal Y},\bar{\cal Y}) = \Phi^{(n)}(Y,\bar{Y}, \Upsilon=\bar \Upsilon = 0)
\label{Fnhol}
\end{equation}
where we used that $Y^I={\cal Y}^I$ for $\Upsilon=0$.
Explicit expressions for $F^{(1)}, F^{(2)}$ and $F^{(3)}$ are given in 
\eqref{expf123} (the normalization used there differs by a factor 
$2i$).

Setting $\Upsilon=0$, one obtains a holomorphic anomaly equation
\begin{equation}
\label{A1}
\frac{\partial F^{(n)}}{\partial \bar{\cal Y}^I} = \frac{1}{2} 
\bar{F}_{I(0)}^{JK} \sum_{r=1}^{n-1} \partial_J F^{(r)} \partial_K
F^{(n-r)} \;,\;\;n\geq 2 \;
\end{equation}
for the functions $F^{(n)}({\cal Y}, \bar{\cal Y})$. This is not the
full anomaly equation for the genus $n$ topological free 
energies of the topological string. The reason is that, so far,
$F_\Upsilon$ and hence $\Phi^{(1)}$ and $F^{(1)}$ are holomorphic, while
for the topological string they are not. This will be addressed in 
the next step where we extend our formalism to the case of 
a non-holomorphic $F_\Upsilon$. For terminological convenience 
we will refer to the functions $F^{(n)}$ as genus $n$ topological
free energies, or free energies for short.\footnote{We remark that
our formalism is independent of an explicit realization by a concrete
topological string model, and in this sense independent of the topological
string. Our formalism is a general framework, for which the topological 
string is one (important) application.}

The hierarachy of equations (\ref{A1}) 
can be re-organised into a master anomaly equation 
for the topological free energy
\[
F_{\rm top}({\cal Y}, \bar{\cal Y}, Q) = \sum_{n=1}^\infty Q^n F^{(n)}({\cal Y},
\bar{\cal Y}) \;,
\]
where the expansion parameter $Q$ is, for the topological string,
related to the topological string coupling. Taking into account 
that $F^{(1)}$ is (so far) holomorphic, it is straightforward 
to verify that (\ref{A1}) follows from 
\begin{equation}
\label{A2}
\frac{\partial F_{\rm top}}{\partial \bar{\cal Y}^I} =
\frac{1}{2} \bar{F}_{I (0)}^{JK} \partial_J F_{\rm top} \partial_K F_{\rm top}
\end{equation}
by expansion in $Q$.

Since $F_\Upsilon$ is the natural master function in the supergravity
formalism, one would like to have a master anomaly equation for 
it.  This is not
straightforward, since the Taylor coefficients of $F_\Upsilon(Y,\Upsilon)$
with respect to $\Upsilon$ are not symplectic functions. We proceed by 
expressing $F_\Upsilon$ in stringy variables and 
introducing a shift in $\Upsilon$: 
\[
F_{\Upsilon} ( {\cal Y}, \bar{\cal Y}, \Upsilon, \bar \Upsilon, Q) :=
F_{\Upsilon} ( {\cal Y}, \bar{\cal Y}, \Upsilon + Q, \bar \Upsilon) \;.
\]
Then we make a Taylor expansion with respect to the `fluctuation' 
$Q$.
As indicated by notation, we need 
to treat $\Upsilon$ and $\bar{\Upsilon}$ as independent
variables, and the `shifted' $F_\Upsilon$ is not any more 
holomorphic in supergravity variables. Also note that when
using supergravity variables the
dependence on $Q$ is not any more of the form $\Upsilon + Q$.

Now we express the expansion coefficients in supergravity variables:
\begin{eqnarray}
F_{\Upsilon} ( {\cal Y}, \bar{\cal Y}, \Upsilon, \bar{\Upsilon}, Q) &=&
\sum_{n=1}^\infty \frac{1}{n!} Q^n \partial^n_\Upsilon F_\Upsilon ({\cal Y}, \bar{\cal Y},
\Upsilon, \bar \Upsilon) =
\sum_{n=1}^\infty \frac{1}{n!} Q^n D^n_\Upsilon F_\Upsilon(Y,\Upsilon) 
\nonumber \\
&=&
\sum_{n=1}^\infty \frac{Q^n}{n!} (n+1)! \Phi^{(n)}(Y,\bar{Y},\Upsilon,
\bar{\Upsilon}) 
\nonumber\\
&=& \sum_{n=1}^\infty (n+1) Q^n  \Phi^{(n)}(Y,\bar{Y},\Upsilon,\bar{\Upsilon}) \;. 
\nonumber
\end{eqnarray}
Next we integrate with respect to $Q$:
\[
G(Y,\bar{Y}, \Upsilon, \bar{\Upsilon}, Q) = \sum_{n=0}^\infty Q^{n+1} \Phi^{(n+1)}(Y,\bar{Y}, 
\Upsilon, \bar{\Upsilon}) \;,
\]
and by setting $\Upsilon=0=\bar{\Upsilon}$ 
we obtain the topological free energy $F_{\rm top}$.
\begin{eqnarray*}
G(Y,\bar{Y}, \Upsilon=\bar{\Upsilon}=0, Q) &=& 
G({\cal Y}, \bar{\cal Y}, \Upsilon=\bar{\Upsilon}=0, Q) =
\sum_{n=0}^\infty Q^{n+1} F^{(n+1)}({\cal Y}, \bar{\cal Y}) \\
& = & F_{\rm top}
({\cal Y}, \bar{\cal Y}, Q) \;.
\end{eqnarray*}
The function $G$ satisfies the master anomaly equation
\[
\frac{\partial} {\partial \bar{Y}^I} G = \frac{1}{2} \bar{F}_I^{JK}
\partial_J G \partial_K G \;,
\]
which for $\Upsilon=\bar{\Upsilon}=0$ becomes the master anomaly equation (\ref{A2})
for the topological free energy.

We note that the relation between the topological free energies
$F^{(g)}$ and the function $F_\Upsilon$ is complicated. This is
of course to be expected from \cite{Cardoso:2014kwa}. 
The reason
is that in a Taylor expansion of $F_\Upsilon(Y,\Upsilon)$ the
coefficients are not symplectic functions. The topological free energies
can be regarded as coefficients in a symplectically covariant 
Taylor expansion, which in practice we cannot manage in closed form
but only by evaluating derivatives at $\Upsilon=0$. We also remark
that the use of two complementary set of complex coordinates, reflects
that there is a tension between holomorphicity and symplecticity. 
In the supergravity variables we have manifest holomorphicity, but 
only the full symplectic vector $(Y^I,F_I)$ and the full function
$F_\Upsilon$ are symplectically covariant. If one wants to organise
data in a hierarchy of symplectic function, holomorphicity is 
violated, albeit in a systematic way controlled by the anomaly
equation. One can then either work with the covariant derivatives
$\Phi^{(n+1)} \simeq D^n_\Upsilon F_\Upsilon$ or use the stringy
variables and work with the free energies $F^{(n)}$.

Above we obtained a master equation for $F_\Upsilon$ from the
master anomaly equation for $F_{\rm top}$. The result is not
quite satisfactory, as we need the background shift $Q$ as
a device. But in the next section we will see that 
a master equation for $F_\Upsilon$ can be obtained directly 
within the supergravity formalism.

\subsection{Anomaly equation from the Hessian structure \label{hessianstr1}}

We will now show that the holomorphic anomaly equation arises
as an integrability condition for the existence of a Hesse
potential on $\hat{M}$.  The metric
$g^H$ being Hessian means that $S=\nabla g^H$ is a completely
symmetric rank three tensor. In $\nabla$-affine coordinates
$Q^a= (x^I, y_I, \Upsilon, \bar{\Upsilon})$ the components
of the tensor $S$ are simply the third partial derivatives
of the Hesse potential, or equivalently the first partial
derivatives of the metric, and therefore proportional to
the Christoffel symbols of the first kind (which for a Hessian
metric are completely symmetric with respect to $\nabla$-affine
coordinates):
\[
S_{abc} = \partial^3_{abc} H = \partial_a g^H_{bc} \;.
\]

One particular relation is
\[
S_{x^I \Upsilon \Upsilon} = S_{\Upsilon x^I \Upsilon} \;,
\]
or
\begin{equation}
\label{HA1}
\partial_{x^I} g^H_{\Upsilon \Upsilon} = \partial_\Upsilon g^H_{x^I \Upsilon}\;,
\end{equation}
where
\[
g^H_{\Upsilon \Upsilon} = - iD_\Upsilon F_\Upsilon 
\]
and 
\[
g^H_{x^I \Upsilon} = 2 \bar{F}_{IJ} N^{JK} F_{K\Upsilon} \;.
\]
We now evaluate equation (\ref{HA1}) 
in supergravity coordinates
$(Y^I, \bar{Y}^I, \Upsilon, \bar{\Upsilon})$, using the 
corresponding Jacobian to obtain
\[
S_{x^I \Upsilon \Upsilon} 
= \left.  \frac{\partial g^H_{\Upsilon \Upsilon}}{\partial {x^I}}\right|_{y} = 
\left.  \frac{\partial g^H_{\Upsilon \Upsilon}}{\partial {x^I}}\right|_{u}
+
\frac{\partial u^K}{\partial x^I} \frac{\partial g^H_{\Upsilon \Upsilon} }{
\partial u^K} \;,\;\;\;\mbox{where}\;\;\;
\left. \frac{\partial}{\partial {x^I}}\right|_{u}
= \frac{\partial}{\partial Y^I} + 
\frac{\partial}{\partial \bar{Y}^I} \;
\]
and 
\[
S_{\Upsilon x^I \Upsilon}  = \left. \frac{\partial g^H_{x^I \Upsilon}}{\partial
\Upsilon}\right|_{x,y} =  \left. \frac{\partial g^H_{x^I \Upsilon}}{\partial
\Upsilon}\right|_{x,u} + \frac{\partial u^K}{\partial \Upsilon} \frac{\partial
g^H_{x^I \Upsilon}}{\partial u^K}  \;.
\]
We find for the various terms,
\begin{eqnarray}
\left. \frac{\partial g^H_{\Upsilon \Upsilon} }{\partial {x^I}}\right|_{u}
& = & -i \frac{\partial}{\partial \bar{Y}^I}
D_\Upsilon F_\Upsilon -i
{\bar F}_{\bar I}{}^{KL} F_{K \Upsilon} F_{L \Upsilon} \;, \nonumber\\
\left. \frac{\partial g^H_{\Upsilon \Upsilon} }{
\partial u^K}\right|_{x} &=&  \left(\frac{\partial}{\partial Y^K} - \frac{\partial}{\partial {\bar Y}^K} \right)
\left( F_{\Upsilon \Upsilon} + i N^{KL} F_{K \Upsilon} F_{L \Upsilon} \right) \nonumber\\
&=& F_{K \Upsilon \Upsilon} - F_K{}^{PQ} F_{P \Upsilon} F_{Q \Upsilon} - 2 N^{PQ} F_{KP \Upsilon} F_{Q \Upsilon}
- {\bar F}_K{}^{PQ} F_{P \Upsilon}  F_{Q \Upsilon} \;, \nonumber\\
\left. \frac{\partial g^H_{x^I \Upsilon}}{\partial
\Upsilon}\right|_{x,u} &=& - i F_{I \Upsilon \Upsilon} + F_{\Upsilon I}{}^J F_{J \Upsilon} + \left(F_{IJ} + {\bar F}_{IJ}
\right)
 \left(i F_{\Upsilon}{}^{JK} F_{K \Upsilon} + F_{\Upsilon \Upsilon}{}^J \right) \;,
\nonumber\\
\left. \frac{\partial
g^H_{x^I \Upsilon}}{\partial u^K}\right|_{x}  &=& F_{IK \Upsilon} + i F_{IK}{}^L F_{L \Upsilon} + i 
\left(F_{IL} + {\bar F}_{IL} \right) \left( i F_K{}^{LP} F_{P \Upsilon} + F_{\Upsilon K}{}^L 
\right)  \nonumber\\
&& - i {\bar F}_{IK}{}^L F_{L \Upsilon} - \left(F_{IL} + {\bar F}_{IL} \right) {\bar F}_{K}{}^{LP} F_{P \Upsilon} 
\;,
\end{eqnarray}
where indices are raised using $N^{IJ}$. Then, the Hessian condition 
(\ref{HA1}) 
results in
\begin{equation}
\label{HA2}
\frac{\partial}{\partial \bar{Y}^I}
D_\Upsilon F_\Upsilon = 
\bar{F}_{{I}{J}{K}} N^{JP} N^{KQ} F_{P\Upsilon}
F_{Q\Upsilon} \;.
\end{equation}
In the holomorphic case at hand ($\partial_{\bar{I}} F_\Upsilon=0$),
this equation can be regarded
as a master anomaly
equation in supergravity variables. First note that 
for $\Upsilon=0$ (\ref{HA2}) reduces to the anomaly equation for $F^{(2)}$.
The anomaly equations for $F^{(n)}$ with $n>2$ are obtained by
covariant differentiation of (\ref{HA2}). 
Here one uses that holomorphicity
of the generalized prepotential implies
\[
\partial_{\bar{I}} F_\Upsilon = 0 \;,\;\;\;
D_\Upsilon \bar{F}_{IJK} = 0 
\]
and one also uses the identity \cite{Cardoso:2012nh}
\[
[D_\Upsilon, N^{IJ}\partial_J]= 0 \;.
\]
For example,
to derive the anomaly equation for $\Phi^{(3)}$ and, hence, for $F^{(3)}$
we need to evaluate
\[
\partial_{\bar{I}} D^2_\Upsilon F_\Upsilon =
D_\Upsilon \partial_{\bar{I}} D_\Upsilon F_\Upsilon + 
i (\partial_{\bar{I}} N^{JK}) F_{J\Upsilon} \partial_K D_\Upsilon F_\Upsilon
\]
\[
= 3 \bar{F}_{I}^{JK} \partial_J F_\Upsilon \partial_K D_\Upsilon
F_\Upsilon \;.
\]
Using that $D^{n-1}_\Upsilon F_\Upsilon = n! \Phi^{(n)}$ this becomes 
\[
\partial_{\bar{I}} \Phi^{(3)} = \bar{F}_{I}^{JK} \partial_J \Phi^{(1)} 
\partial_K \Phi^{(2)} = \frac{1}{2}  \bar{F}_{I}^{JK}  
\sum_{r=1}^2 \partial_J \Phi^{(r)} 
\partial_K \Phi^{(3-r)}  \;,
\]
which for $\Upsilon=0$ is the anomaly equation for $F^{(3)}$. 
Proceeding by induction one obtains the full hierarchy
(\ref{A0}) of anomaly equations.

One may ask whether other components of $S$ will give rise to additional non-trivial
differential equations. To investigate this, we now consider the component $S_{x^I \Upsilon \bar \Upsilon}
= \partial_{x^I} g^H_{\Upsilon \bar \Upsilon}$,
which is constructed out of the 
metric component
$g^H_{\Upsilon \bar \Upsilon} = N^{IJ} F_{I \Upsilon} {\bar F}_{J \Upsilon}$.
Evaluating the relation 
$S_{x^I \Upsilon  \bar \Upsilon} = S_{\bar \Upsilon x^I \Upsilon} = \partial_{\bar \Upsilon} \,  g^H_{x^I \Upsilon}$
in supergravity variables we find that it is identically satisfied. Thus, the only non-trivial differential equation resulting from 
$g^H_{\Upsilon \Upsilon}$ and $g^H_{\Upsilon \bar \Upsilon}$ is encoded in the relation $S_{x^I \Upsilon \Upsilon} = S_{\Upsilon x^I \Upsilon}$.

\section{The non-holomorphic deformation \label{nholskg}}

So far we have assumed that $F(Y,\Upsilon)$ and, hence, $F_\Upsilon$
are holomorphic in the supergravity variables, which implies
that $F^{(1)}({\cal Y})$ is also holomorphic, while $F^{(g)}({\cal Y},
\bar{\cal Y})$ with $g>1$ satisfy the anomaly equation (\ref{A1}). 
For the topological string the situation is more complicated since
already $F^{(1)}({\cal Y}, \bar{\cal Y})$ is non-holomorphic. We
therefore now generalize our framework and induce geometric data
on $\hat{M}$ using a non-holomorphic map $\phi: \hat{M} \rightarrow V$,
which then corresponds to a non-holomorphic generalized prepotential
$F=F(Y,\bar{Y}, \Upsilon, \bar{\Upsilon})$. This explicit non-holomorphicity
will in turn modify the anomaly equation (\ref{A1}) satisfied by 
the topological free energies $F^{(g)}({\cal Y}, \bar{\cal Y})$. 
The precise form of the modification depends on the details of the
non-holomorphic deformation. We will first keep the discussion general,
and later show that when chosing a particular non-holomorphic deformation
we obtain the correct full
anomaly equation (at least to leading order in a formal expansion
we will explain later). As discussed in \cite{Cardoso:2012nh} there
are other types of non-holomorphic deformations that are, for example,
relevant for non-linear deformations of electrodynamics. Any such 
deformation could be analyzed in the framework of our formalism.

Since $F$ and $F_\Upsilon$ are no longer holomorphic, they will have 
non-vanishing derivatives with respect to $\bar{Y}^I$ and $\bar{\Upsilon}$.
In the following we will use a notation involving `unbarred' indices
$I,J,\ldots $ and `barred' indices $\bar{I}, \bar{J}, \ldots$.

\subsection{Non-holomorphic deformation of the prepotential}

We generalize the map (\ref{phi_holo})
to 
\begin{equation}
\label{phi_non_holo}
\phi\;: \hat{M} = M \times \mathbbm{C} \rightarrow V \;,\;\;\;
(Y^I, \Upsilon) \mapsto (Y^I, F_I(Y,\bar{Y},\Upsilon,\bar{\Upsilon})) \;,
\end{equation}
where $F_I = \partial F/\partial Y^I$, can be obtained from a 
generalized prepotential $F$. We assume that $F$ 
has the form \cite{LopesCardoso:2006bg}
\begin{equation}
\label{gF=F+Omega}
F(Y,\bar{Y}, \Upsilon, \bar{\Upsilon}) = 
F^{(0)}(Y) + 2 i \Omega(Y,\bar{Y}, \Upsilon, \bar{\Upsilon}) \;,
\end{equation}
where $F^{(0)}$ is the undeformed prepotential, and 
where $\Omega$ is real-valued.\footnote{This function is not
to be confused with the complex symplectic form on the vector
space $V$ introduced in subsection 2.1.}
 The holomorphic deformation is recovered
when $\Omega$ is harmonic. This makes use of the observation 
that the complex symplectic vector $(Y^I, F_I)$ does not uniquely 
determine the prepotential $F$ \cite{Cardoso:2014kwa}. 
If we make a transformation 
\begin{eqnarray*}
F^{(0)}(Y) &\rightarrow&  F^{(0)}(Y) + g (Y,\Upsilon) \;,\;\;\; \\
\Omega(Y,\bar{Y}, \Upsilon, \bar{\Upsilon}) &\rightarrow & 
\Omega(Y,\bar{Y}, \Upsilon, \bar{\Upsilon}) - \frac{1}{2i}
(g(Y,\Upsilon) - \bar{g}(\bar{Y}, \bar{\Upsilon}))  \;,
\end{eqnarray*}
where $g(Y,\Upsilon)$ is holomorphic, then $F$
changes by an antiholomorphic function, $F\rightarrow F + \bar{g}$,
and the symplectic vector $(Y^I,F_I)$ and the map $\phi$ are 
invariant. If $\Omega$ is harmonic, 
\[
\Omega(Y,\bar{Y},\Upsilon,\bar{\Upsilon}) = f(Y,\Upsilon) 
+ \bar{f} (\bar{Y}, \bar{\Upsilon}) 
\;,
\]
we can make a transformation with $g =  2i f$ and obtain
\[
F \rightarrow F^{(0)}(Y) + 2 i f(Y,\Upsilon) =: F(Y,\Upsilon)\;,
\]
which is a holomorphically deformed prepotential, as considered in the
previous section. If, however, $\Omega$ is not harmonic, then we have
a genuine generalization which requires us to consider non-holomorphic
generalized prepotentials.
For the case of the topological string, it is convenient to 
split the non-holomorphic generalized prepotential as in 
(\ref{gF=F+Omega}) into the undeformed prepotential $F^{(0)}$ 
and a real-valued non-harmonic function $\Omega$ which encodes
all higher derivative effects, holomorphic as well as non-holomorphic.

We proceed by analysing the geometry induced by pulling back 
the standard hermitian form $\gamma_V$ of $V$ given by (\ref{gammaV})
to $\hat{M}$ using (\ref{phi_non_holo}):  
\[
\gamma= - i (F^{(0)}_{IJ} - \bar{F}^{(0)}_{\bar{I}\bar{J}}) dY^I \otimes
d \bar{Y}^J 
+ 2 (\Omega_{IJ} + \Omega_{\bar{I} \bar{J}}) dY^I \otimes d \bar{Y}^J
+ 2 \Omega_{\bar{I}J} dY^I \otimes dY^J 
\]
\[
+ 2 \Omega_{I \bar{ J}} d\bar{Y}^I 
\otimes d \bar{Y}^J 
+ 2 \Omega_{\bar{I} \bar{\Upsilon}} dY^I \otimes d \bar{\Upsilon}
+ 2 \Omega_{I \Upsilon} d \Upsilon \otimes d \bar{Y}^I
+ 2 \Omega_{\bar{I}\Upsilon} dY^I \otimes d\Upsilon 
\]
\[
+ 2 \Omega_{I \bar{\Upsilon}} d \bar{\Upsilon} \otimes d \bar{Y}^I\;.
\]
By decomposing $\gamma = g + i \omega$, we obtain the following 
metric on $\hat{M}$:
\[
g = - i (F^{(0)}_{IJ} - \bar{F}^{(0)}_{\bar{I}\bar{J}}) dY^I d\bar{Y}^J
+ 2 (\Omega_{IJ} + \Omega_{\bar{I} \bar{J}}) dY^I d \bar{Y}^J
+ 2 \Omega_{\bar{I}J} dY^I dY^J + 2 \Omega_{I\bar{J}} d\bar{Y}^I  d \bar{Y}^J 
\]
\[
+ 2 \Omega_{\bar{I} \bar{\Upsilon}} dY^I  d \bar{\Upsilon}
+ 2 \Omega_{I \Upsilon} d \Upsilon  d \bar{Y}^I
+ 2 \Omega_{\bar{I}\Upsilon} dY^I  d\Upsilon 
+ 2 \Omega_{I \bar{\Upsilon}} d \bar{\Upsilon}  d \bar{Y}^I \;.
\]
From this expression it is manifest that $g$ is not Hermitian, and
hence not K\"ahler with respect to the natural complex structure $J$.
The non-Hermiticity is encoded in the mixed derivatives $\Omega_{I\bar{J}}$,
which makes it manifest that it is related to the non-harmonicity
of $\Omega$. This metric occurs in the sigma model discussed
in \cite{Cardoso:2012mc}. 

The imaginary part of $\gamma$ defines a two-form on $\hat{M}$: 
\begin{eqnarray}
\omega &=& \frac{1}{2i} (- i (F^{(0)}_{IJ} - \bar{F}^{(0)}_{\bar{I} \bar{J}}))
dY^I \wedge d\bar{Y}^J - i (\Omega_{\bar{I} \bar{J}}  + \Omega_{IJ})
dY^I \wedge d\bar{Y}^J  \nonumber \\
&& - i \Omega_{\bar{I} J} dY^I \wedge d Y^J 
+ i \Omega_{I\bar{J}} d\bar{Y}^I \wedge d \bar{Y}^J 
- i \Omega_{\bar{I}\bar{\Upsilon}} dY^I \wedge d \bar{\Upsilon}
- i \Omega_{I \Upsilon} d \Upsilon \wedge d\bar{Y}^I 
\nonumber \\
& & 
- i \Omega_{\bar{I} \Upsilon} dY^I \wedge d \Upsilon
+ i \Omega_{I\bar{\Upsilon}} d\bar{Y}^I \wedge d \bar{\Upsilon} \;.
\label{omega_non_holo}
\end{eqnarray}
This two-form is no longer of type $(1,1)$ with respect to the 
standard complex structure, which is consistent with the
non-Hermiticity of $g$. However, $\omega$ is still closed
\[
d\omega=0 \;,
\]
so that $(\hat{M},\omega)$ is at least a symplectic manifold. 

Comparing $\omega$ to $dx^I \wedge dy^I$,  we find: 
\begin{equation}
\label{dxdy_non_holo}
2 dx^I \wedge dy_I = \omega + 2 i   \Omega_{I\bar{J}} 
dY^I \wedge d\bar{Y}^J 
+ i \Omega_{I\Upsilon} d Y^I \wedge d\Upsilon 
- i \Omega_{\bar{I} \bar{\Upsilon}} d\bar{Y}^I \wedge d\bar{\Upsilon}
\end{equation}
\[
+ i \Omega_{I \bar{\Upsilon}} dY^I \wedge d \bar{\Upsilon}
- i \Omega_{\bar{I} \bar{\Upsilon}} d \bar{Y}^I \wedge d\Upsilon \;.
\]
As a consistency check, we verify that $2 dx^I \wedge dy_I - \omega$
is closed:
\[
2 dx^I \wedge dy_I  = \omega - \frac{1}{2} d (\Upsilon d F_\Upsilon)
- \frac{1}{2} d ( \bar{\Upsilon} d F_{\bar{\Upsilon}}) 
- \partial \overline{\partial} F \;,
\]
where $\partial = dY^I \otimes \partial_{Y^I} + d \Upsilon \otimes 
\partial_\Upsilon$. Note that the difference between the symplectic
form $2dx^I \wedge dy_I$ of $M$ and the symplectic form $\omega$ 
of $\hat{M}$ is still exact. Compared to (\ref{Diff_holo}) we have 
an additional term which measures the non-holomorphicity of the generalized 
prepotential.

\subsection{Real coordinates and the Hesse potential \label{realHe}}

To convert from complex supergravity variables we need to generalize
our previous calculation of the Jacobian and its inverse: 
\[
\frac{D(x,y,\Upsilon, \bar{\Upsilon})}{D(x,u,\Upsilon,\bar{\Upsilon})} =
\left( \begin{array}{cccc}
\mathbbm{1} & 0 & 0 & 0 \\
\frac{1}{2} R_+ & - \frac{1}{2} N_- & \frac{1}{2} (F_{I\Upsilon} + 
\bar{F}_{\bar{I}\Upsilon}) & \frac{1}{2} ( \bar{F}_{\bar{I} \bar{\Upsilon}}
+ F_{I\bar{\Upsilon}}) \\
0 & 0 & \mathbbm{1} & 0 \\
0 & 0 & 0 & \mathbbm{1} \\
\end{array} \right)
\]
and, by a straigtforward matrix inversion
\[
\frac{D(x,u,\Upsilon,\bar{\Upsilon})}{D(x,y,\Upsilon, \bar{\Upsilon})} =
\left( \begin{array}{cccc}
\mathbbm{1} & 0 & 0 & 0 \\
N_-^{-1}  R_+ & - 2 N_-^{-1}  & N_-^{-1} (F_{I\Upsilon} + 
\bar{F}_{\bar{I}\Upsilon}) & N_-^{-1} ( \bar{F}_{\bar{I} \bar{\Upsilon}}
+ F_{I\bar{\Upsilon}}) \\
0 & 0 & \mathbbm{1} & 0 \\
0 & 0 & 0 & \mathbbm{1} \\
\end{array} \right) \;.
\]
This reduces to the previous result when switching off the non-holomorphic
deformation. When restricting to the left upper block, the result agrees
with \cite{Cardoso:2012mc}. We have used the following definitions
\cite{Cardoso:2012mc}:
\[
N_{\pm IJ} = N_{IJ} \pm 2 \mbox{Im} F_{I\bar{J}} = - i (F_{IJ} - \bar{F}_{\bar{I}
\bar{J}} \pm  F_{I\bar{J}} \mp \bar{F}_{\bar{I} J} )
\]
and 
\[
R_{\pm IJ} = R_{IJ}  \pm 2 \mbox{Re} F_{I\bar{J}} = 
F_{IJ} + \bar{F}_{\bar{I} \bar{J}} \pm F_{I\bar{J}} \pm \bar{F}_{\bar{I} J} \;.
\]
Note that $N_-^T = N_-$, while $R_{\pm}^T = R_{\mp}$.

As already observed in \cite{LopesCardoso:2006bg}, 
in the presence of 
explicit non-holomorphic deformations the Hesse potential
is not to be defined as the Legendre transform of $2 \mbox{Im} F$ 
but rather as the Legendre of 
\begin{equation}
\label{lforh}
L = 2 \mbox{Im} F - 2 \Omega = 2 \mbox{Im} F^{(0)} + 2 \Omega \;.
\end{equation}
As explained
in \cite{Cardoso:2012nh,Cardoso:2014kwa}, the function $L$ can be interpreted as a 
Lagrange function, and the Hesse potential as the corresponding 
Hamilton function.\footnote{In 
\cite{Cardoso:2012nh,Cardoso:2014kwa} the Hesse potential is
normalized differently by a factor 2 compared to \cite{LopesCardoso:2006bg}
and the present paper.}
Thus the Hesse potential associated to a non-holomorphically deformed
prepotential is
\[
H(x,y,\Upsilon,\bar{\Upsilon}) = 
-i (F - \bar{F}) - 2 \Omega 
- 2 u^I y_I 
\;.
\]
By taking derivatives with respect to the real coordinates
$(Q^A)=(q^a, \Upsilon, \bar{\Upsilon})$, where $(q^a)=(x^I, y_I)$ 
we obtain the components of a Hessian metric:
\[
\frac{\partial H}{\partial q^a \partial q^b} =
\left( 
\begin{array}{cc}
N_+  + R_- N_-^{-1} R_+  & - 2 R_- N_-^{-1}  \\
- 2 N_-^{-1} R_+ & 4 N_-^{-1} \\
\end{array} \right) \;,
\]
\[
\frac{\partial^2 H}{\partial x^I \partial \Upsilon} =
- i (F_{I\Upsilon} - \bar{F}_{\bar{I} \Upsilon})
+ R_{-IK} N_-^{KJ} ( F_{J\Upsilon} + \bar{F}_{\bar{J} \Upsilon})
\;,\;\;\;
\frac{\partial^2 H}{\partial y_I \partial \Upsilon} = 
- 2 N_-^{IK} (F_{K\Upsilon} + \bar{F}_{\bar{K} \Upsilon}) \;,
\]
together with their complex conjugates and
\[
\frac{\partial^2 H}{\partial \Upsilon \partial \bar{\Upsilon}} =
- i F_{\Upsilon \bar{\Upsilon}} + N_-^{IJ} ( \bar{F}_{\bar{I} \bar{\Upsilon}}
- \bar{F}_{I \bar{\Upsilon}})(F_{\Upsilon J} - F_{\Upsilon \bar{J}})
= - i D_{\Upsilon} F_{\bar{\Upsilon}}  \;,\;\;\;
\]
\[
\frac{\partial^2 H}{\partial \Upsilon \partial \Upsilon} = 
- i D_{\Upsilon} F_\Upsilon \;,\;\;\;
\frac{\partial^2 H}{\partial \bar{\Upsilon} \partial \bar{\Upsilon}} = 
i \overline{D_{\Upsilon} F_\Upsilon} \;,\;\;\;
\]
where 
\begin{equation}
D_\Upsilon = \partial_\Upsilon + i {N}_-^{IJ} ( F_{\Upsilon J} - 
F_{\Upsilon \bar{J}} ) \left( \frac{\partial}{\partial Y^I} -
\frac{\partial}{\partial \bar{Y}^I} \right)
\label{D_Upsilon_non_holo}
\end{equation}
is the symplectic covariant derivative introduced in 
\cite{Cardoso:2012nh}. This covariant derivative is a 
generalization of (\ref{D_Upsilon_holo}) which generates 
a hierarchy of symplectic functions starting from a 
non-holomorphic symplectic function. For holomorphic symplectic
functions it reduces to (\ref{D_Upsilon_holo}). We will show
below that $D_\Upsilon$ can be derived by transforming the
partial derivative $\left. \partial_\Upsilon \right|_{{\cal Y}}$ 
from stringy variables to supergravity variables.

As before, the Hessian metric $g^H$ differs from the metric $g$
induced by pulling back $g_V$ using $\phi$ by differentials involving
derivatives of $H$ with respect to $\Upsilon, \bar{\Upsilon}$:
\[
g^H = g + \left. \partial^2H \right|_{x,y}
\]
where 
\[
\left. \partial^2H \right|_{x,y} = 
\frac{\partial^2 H}{\partial \Upsilon \partial \Upsilon} d\Upsilon d\Upsilon +
2 \frac{\partial^2 H}{\partial \Upsilon \partial \bar{\Upsilon}} d\Upsilon 
d\bar{\Upsilon} +
\frac{\partial^2 H}{\partial \bar{\Upsilon} \partial \bar{\Upsilon}} 
d\bar{\Upsilon} d\bar{\Upsilon}  \;.
\]


\subsection{The symplectic covariant derivative}

As before we can use (\ref{omega_non_holo})
and (\ref{dxdy_non_holo}) to obtain exact information about
the coordinate transformation $\Delta Y^I = {\cal Y}^I - Y^I$ 
between supergravity variables and stringy variables. By proceeding 
as in subsection \ref{scc}, namely
converting the expression (\ref{dxdy_non_holo}) from supergravity
variables to stringy variables, and then converting the result
to real variables, we find the following consistency condition:
\begin{equation}
\label{Delta_non_holo}
\frac{\partial \Delta Y^I}{\partial \Upsilon} =
- i \hat{N}^{IK} (F_{K \Upsilon} + \bar{F}_{\bar{K} \Upsilon}) \;,
\end{equation}
where $\hat{N}^{IJ}$ is the inverse of the matrix
\[
\hat{N}_{IJ} = 
N_{IJ} + i F_{J\bar{I}} - i \bar{F}_{\bar{J}I}
= - i ( F_{IJ} - \bar{F}_{\bar{I}\bar{J}} - F_{\bar{I}J}
+ \bar{F}_{I \bar{J}} ) \]
\[
=
- i (F_{IJ} - \bar{F}_{\bar{I} \bar{J}} - 2i \Omega_{J\bar{I}}
- 2 i \Omega_{\bar{J}I})
\]
defined in \cite{Cardoso:2012nh}. 
Note that $\hat{N}_{IJ} = N_{-IJ}$, which was defined before.
The formula (\ref{Delta_non_holo})
can be used to derive a modified symplectically covariant derivative,
which allows to generate new symplectic functions given a 
non-holomorphic symplectic function $G(Y,\bar{Y},\Upsilon,\bar{\Upsilon})$.
Indeed, if such a function is given we can express it in stringy
variables, $G=G(Y({\cal Y}, \bar{\cal Y}, \Upsilon,\bar{\Upsilon}),
\bar{Y}({\cal Y}, \bar{\cal Y}, \Upsilon, \bar{\Upsilon}), \Upsilon,
\bar{\Upsilon})$, and we know that $\left. \partial G/\partial \Upsilon 
\right|_{{\cal Y}}$ is a symplectic function. Expressing this function
in supergravity variables we obtain
\[
\left. \frac{\partial}{\partial \Upsilon} \right|_{ {\cal Y }} G =
\left. \frac{\partial}{\partial \Upsilon}\right|_{Y} G -
\frac{\partial \Delta Y^I}{\partial \Upsilon} (\partial_{Y^I} 
- \partial_{\bar{Y}^I} ) G =: 
D_\Upsilon G \;,
\]
where $D_\Upsilon$ denotes the symplectically covariant derivative introduced in \eqref{D_Upsilon_non_holo}, as 
can be readily verified by using $F_\Upsilon = 2 i \Omega_\Upsilon$, where
$\Omega$ is real valued, which implies
$\bar{F}_{\bar{I}\Upsilon} = - F_{\Upsilon \bar{I}}$.
This derivative operator
was already found in \cite{Cardoso:2012nh} 
based on studying the symplectic transformation of derivatives of
a non-holomorphic generalized prepotential. We have now derived
this covariant derivative from a coordinate transformation. 

The symplectically covariant derivative $D_\Upsilon$  can be applied 
to any non-holomorphic symplectic function. Thus,
we can now construct a hierarchy of symplectic functions
starting from a non-holomorphic $F_\Upsilon$: 
\begin{equation}
\Phi^{(n+1)} = \frac{1}{(n+1)!} D^n_\Upsilon F_\Upsilon (Y,\bar{Y},\Upsilon, 
\bar{\Upsilon}) \;.
\label{Phihier}
\end{equation}
As before, we define topological free
energies by
\[
F^{(n)}({\cal Y}, \bar{\cal Y}) := 
\left. \Phi^{(n)}(Y,\bar{Y},\Upsilon,\bar{\Upsilon}) \right|_{\Upsilon=0} 
\;,\;\;\; n\geq 2 \;.
\]
These functions will satisfy a holomorphic anomaly
equation, whose precise form depends on the details of the 
non-holomorphic deformation.

\subsection{The holomorphic  anomaly equation}

We would like to show that for a suitable choice of a non-holomorphic
deformation we obtain the holomorphic anomaly equation of the topological string. 
As anticipated from \cite{Cardoso:2014kwa} this is laborious to 
do explicitly, since an explicit non-holomorphic deformation leads
to a proliferation of non-holomorphic terms. As in \cite{Cardoso:2014kwa}
we will resort to a (formal) series expansion in parameters which control
the non-holomorphicity, and rely on results about the symplectic
transformation behaviour of various quantites.

The non-holomorphic dependence of the topological free energies $F^{(g)}$ is entirely
encoded in $N^{IJ}_{(0)}$ (which, we recall, is the inverse of $N_{IJ}^{(0)} = -i ( F_{IJ}^{(0)} - {\bar F}_{IJ}^{(0)} )$).
The higher $F^{(g)}$ (with $g \geq 2$) are polynomials of degree $3g-3$ in 
$N^{IJ}_{(0)}$, while $F^{(1)}$ depends on the logarithm of $\det N^{IJ}_{(0)}$.
In the following, we will focus on the polynomial dependence on $N_{IJ}^{(0)}$
of the higher $F^{(g)}$, keeping $F^{(1)}$ holomorphic for the time being.  We thus consider the deformation
\begin{equation}
\Omega(Y, \bar Y, \Upsilon, \bar \Upsilon) = f(Y, \Upsilon) + 2 \beta \, \Upsilon \, N^{IJ}_{(0)} \, f_{IJ}(Y, \Upsilon) + {\rm c.c}  \;.,
\label{def-beta1}
\end{equation}
where
the departure from harmonicity is encoded in $ N^{IJ}_{(0)}$. 
We will work to first order in the deformation
parameter $\beta$ and in $ N^{IJ}_{(0)}$ to avoid a proliferation of new terms 
compared to the holomorphic case. Note that in applications $\beta$ and 
$N^{IJ}_{(0)}$ are not necessarily small, so that the expansion is formal.
The above defines a toy model that, as we will see, reproduces
the holomorphic anomaly equation for the topological free energies
$F^{(g)}$ for $g \geq 2$, to leading order in  $\beta$ and in $ N^{IJ}_{(0)}$.
Since $F^{(1)}$ is still holomorphic, this toy model does not fully capture
the topological string. We will address this issue at the end of this section.

Expanding
\begin{equation}
f(Y,\Upsilon) = \sum_{n=1}^\infty \Upsilon^n \, f^{(n)}(Y)
\Rightarrow
f_\Upsilon = \sum _{n=1}^\infty n \Upsilon^{n-1} f^{(n)}(Y) \;,
\label{serom}
\end{equation}
we obtain from \eqref{def-beta1}\,
\begin{equation}
\Omega_{\Upsilon} = f^{(1)} + \sum _{n=2}^\infty n \Upsilon^{n-1} 
\left(f^{(n)}(Y) + 2 \beta \, N_{(0)}^{IJ} \, f^{(n-1)}_{IJ} \right) \;.
\end{equation}
To first order in $\beta$ and in $N^{IJ}_{(0)}$, the function
$F_{\Upsilon} = 2 i \Omega_{\Upsilon}$, given by 
\begin{eqnarray}
\label{symlfn}
F_\Upsilon (Y,\bar{Y},\Upsilon) &=&
2 i \left[f_{\Upsilon}  + 
2 \beta \, N^{IJ}_{(0)} \, f_{IJ}(Y, \Upsilon)
+
2 \beta \, \Upsilon \, N^{IJ}_{(0)} \, f_{\Upsilon IJ}(Y, \Upsilon)
\right] \\
&=&
2 i \left[
f^{(1)}(Y)  +
2 \Upsilon \left( f^{(2)}(Y) + 2 \beta \, N^{IJ}_{(0)} \, f^{(1)}_{IJ}(Y) \right) 
\right.
\nonumber\\
&&
\left. 
\hskip 25mm
+ 3 \Upsilon^2 \left(f^{(3)}(Y) + 2 \beta \, N^{IJ}_{(0)} \, f^{(2)}_{IJ}(Y) \right)
+ \cdots \right] \;, \nonumber
\end{eqnarray}
transforms as a function under symplectic transformations provided we modify the transformation behaviour of $f({Y}, \Upsilon)$ to
(note that we are using supergravity coordinates ${Y}^I$),
\begin{equation}
f ({Y}, \Upsilon) \longrightarrow f({Y}, \Upsilon) + 2 i  \beta  \, \Upsilon \,  {\cal Z}_{0}^{IJ}  \, f_{IJ} (Y, \Upsilon) 
  \;,
\label{transflawom}
\end{equation}
to first order in $\beta$ and in ${\cal Z}^{IJ}_0$. 
Here, ${\cal Z}^{IJ}_{0}$ denotes the transformation matrix
given in \eqref{SZ}.
Note that $ \beta \, N^{IJ}_{(0)} \, f_{IJ}$ transforms as follows \cite{Cardoso:2014kwa}
under symplectic transformations, 
to first order in $\beta$ and in $N^{IJ}_{(0)}$ (or 
${\cal Z}^{IJ}_0$), 
\begin{eqnarray}
\beta \, N^{IJ}_{(0)} \, f_{IJ} &\longrightarrow & \beta \, \left(N^{IJ}_{(0)} - i {\cal Z}^{IJ}_{0} \right) \left(
f_{IJ} - F^{(0)}_{IJL} {\cal Z}^{LP}_{0} f_P \right) \nonumber\\
&=& \beta \, \left(N^{IJ}_{(0)} - i {\cal Z}^{IJ}_{0} \right) 
f_{IJ} + {\cal O}(N^{-1} {\cal Z}_{0}, {\cal Z}_{0}^2) \:.
\label{transwdd}
\end{eqnarray}
Using \eqref{transflawom} and \eqref{transwdd}, it follows that \eqref{symlfn} is a symplectic function
at this order.

We now observe that
\begin{equation}
F_{{\bar I} \Upsilon} \propto \beta \, \partial_{\bar I} N^{JK}_{(0)} \, 
\end{equation}
which is of higher order in $N^{-1}_{(0)}$, and hence will be dropped. Thus
\[
\left. D^n_\Upsilon F_\Upsilon \right|_{\Upsilon=0}  =
\left. D^n_{(0)\Upsilon} F_\Upsilon \right|_{\Upsilon=0}  
+ O ( (N^{-1}_{(0)})^2) \;,
\]
where by 
\[
D_{(0)\Upsilon} = \frac{\partial}{\partial \Upsilon}
+ i N^{IJ} F_{I\Upsilon} \frac{\partial}{\partial Y^J}
\]
we denote the symplectically 
covariant derivative (\ref{D_Upsilon_holo}). Thus, while when 
starting with a non-holomorphic function (\ref{D_Upsilon_holo})
must normally be replaced by (\ref{D_Upsilon_non_holo}), we 
neglect the additional terms in the present context because 
they will necessarily bring in higher powers of $N^{-1}_{(0)}$.

Hence, by working to order $\beta$ and 
neglecting terms of order $(N^{-1}_{(0)})^2$, we obtain 
\begin{eqnarray}
\left. 
D^n_{\Upsilon} F_\Upsilon
\right|_{\Upsilon =0 } &=&
\left . \left. D^n_{(0)\Upsilon} ( 2i f_\Upsilon ) 
\right|_{\Upsilon =0 } 
+ 4 i \beta N^{IJ}_{(0)} \partial^n_\Upsilon \left(
\sum_{m=1}^\infty (m+1)  \Upsilon^{m} f^{(m)}_{IJ} 
\right)\right|_{\Upsilon=0}  \nonumber\\
&=& 
\left. D^n_{(0)\Upsilon} ( 2i f_\Upsilon ) 
\right|_{\Upsilon =0 } 
+ 4 i \beta \,   (n+1)! \, N^{IJ}_{(0)} \, f^{(n)}_{IJ} \;.
\end{eqnarray}
Now we consider the hierarchy 
\eqref{Phihier},
\begin{eqnarray*}
F^{(n+1)} &=&\left.  \frac{D^n_{\Upsilon} F_\Upsilon}{
(n+1)!}\right|_{\Upsilon=0} = \frac{1}{(n+1)!} \left. D^n_{(0)\Upsilon} (2i f_\Upsilon)
\right|_{\Upsilon=0} + 2 \beta \,  N^{IJ}_{(0)} \left. (2i f^{(n)}_{IJ})
\right|_{\Upsilon=0} \\ & & + O ( \beta^2, (N_{(0)}^{-1})^2) \;.
\end{eqnarray*}
For $\beta=0$ only the first term $F^{(n+1)}_{\rm holo} =
[(n+1)!]^{-1} D^n_{(0)\Upsilon} \left. (2i f_\Upsilon)\right|_{\Upsilon=0}$  is present, which
satisfies the anomaly equation (\ref{A1}). Including the deformation 
term of order $\beta$ we get
 (note that $Y^I = {\cal Y}^I $ when $\Upsilon =0$) 
\[
\frac{\partial}{\partial \bar{\cal Y}^K} F^{(n+1)} =
\frac{\partial}{\partial \bar{\cal{Y}}^K} F^{(n+1)}_{\rm holo} 
- 2i  \beta \,  \bar{F}_K^{(0)IJ} \left. (2i f^{(n)}_{IJ}) \right|_{\Upsilon=0}
\]
\[
= \frac{1}{2} \bar{F}_K^{(0) IJ} \left( \sum_{r=1}^n \partial_I F^{(r)}_{\rm holo}
\partial_J F^{(n+1-r)}_{\rm holo} 
- 4 i \beta \,  F^{(n)}_{{(\rm holo)} I J}  \right) \;.
\]
Redefining $F^{(n)} \rightarrow 2i F^{(n)}$ (with $n \geq 1$), we obtain
\begin{equation}
\frac{\partial}{\partial \bar{\cal Y}^K} F^{(n+1)} 
= i F^{(0)IJ}_K \left( \sum_{r=1}^n \partial_I F^{(r)} 
\partial_J F^{(n+1-r)} - 2 \beta D_I \partial_J F^{(n)} \right) \;, 
\label{fullhaeg2}
\end{equation}
up to terms of higher order in $N^{-1}_{(0)}$. Here we used
that the Levi-Civita connection $D_I$ and the non-holomorphic deformation
terms of the $F^{(n)}$ involve at least one further power of
$N^{-1}_{(0)}$, which can be dropped at the order we are working at.
Note that \eqref{fullhaeg2} is the full
holomorphic anomaly equation for the higher $F^{(n)}$ in big moduli space
\cite{Cardoso:2014kwa}. The standard normalization of the anomaly equation is obtained by 
setting $- 2 \beta =1$.

Let us return to the transformation law \eqref{transflawom}.
Inserting the expansion \eqref{serom} into it, we see that $f^{(1)}$ remains invariant
under symplectic transformations. This is not what happens in topological string theory,
where $f^{(1)}$ transforms into $f^{(1)}\rightarrow f^{(1)} + \alpha \, \ln \det
{\cal S}_0$, with $\alpha \in \mathbb{R}$. This transformation behavior is, in turn, compensated
for by the presence of an additional term  $\ln \det N^{(0)}_{IJ}$, which ensures that the topological
free energy $F^{(1)}$, given by 
\[ F^{(1)} ({\cal Y}, \bar{\cal Y}) = 2i \left( f^{(1)} ({\cal Y}) + {\bar f}^{(1)} ({\bar{\cal Y}}) + 
\alpha \ln \det N^{(0)}_{IJ} \right) \;,
\]
is invariant under symplectic transformations. If we now insist that $F^{(1)}$ and $\Omega$ are related
by $F^{(1)} = 2i \Omega_{\Upsilon}|_{\Upsilon =0}$, then this is only possible if we take $\Upsilon$ to be real,
in which case $\Omega$ given in 
\eqref{def-beta1} gets modified to
\begin{equation}
\Omega(Y, \bar Y, \Upsilon) = \left( f(Y, \Upsilon) + 2 \beta \, \Upsilon \, N^{IJ}_{(0)} \, f_{IJ}(Y, \Upsilon) + {\rm c.c.} 
\right) + \alpha \Upsilon \ln \det N^{(0)}_{IJ} \;,
\label{omalbe}
\end{equation}
to first order in $\alpha$. 
Thus, while the deformation $\beta$
 did not enforce any restriction on $\Upsilon$, the presence of the deformation $\alpha$ does.

So far, we restricted ourselves to working at first order in $\alpha$, $\beta$ and $N^{-1}_{(0)}$.
At higher order, the analysis in \cite{Cardoso:2014kwa} shows that $\alpha$ and $\beta$ get locked onto the
same value $\alpha = \beta$. This is a consequence of the requirement that $\Omega$ should transform consistently
under symplectic transformations.
In this way we recover the non-holomorphic deformation relevant for the topological 
string. 

\subsection{From Hessian structure to the full anomaly equation }

We now show how to recover the holomorphic anomaly equation \eqref{fullhaeg2} from the underlying Hessian structure.
We proceed as in subsection \ref{hessianstr1}, and consider the 
totally symmetric
rank three tensor $S= \nabla g^H$, where $g^H$ denotes the Hessian metric computed in subsection \ref{realHe}.
As before, we consider the components
\[
S_{x^I \Upsilon \Upsilon} = S_{\Upsilon x^I \Upsilon} \;.
\]
Since $S$ is a tensor, we can evaluate this relation in other 
coordinate systems, in particular in complex supergravity coordinates. 
As before, using
\[
g^H_{\Upsilon \Upsilon} = - i D_\Upsilon F_\Upsilon \;,
\]
\[
S_{x^I \Upsilon \Upsilon} = \left. \partial_{x^I} \right|_{y}  g^H_{\Upsilon \Upsilon} = 
\left. \partial_{x^I} \right|_{u}  g^H_{\Upsilon \Upsilon} +
\frac{\partial u^K}{\partial x^I} \frac{\partial g^H_{\Upsilon \Upsilon} }{
\partial u^K} \;,\;\;\;\mbox{where}\;\;\;
\left. \partial_{x^I} \right|_{u} = \frac{\partial}{\partial Y^I} + 
\frac{\partial}{\partial \bar{Y}^I} \;,
\]
\[
S_{\Upsilon x^I \Upsilon}  = \left. \frac{\partial g^H_{x^I \Upsilon}}{\partial
\Upsilon}\right|_{x,y} =  \left. \frac{\partial g^H_{x^I \Upsilon}}{\partial
\Upsilon}\right|_{x,u} + \frac{\partial u^K}{\partial x^I} \frac{\partial
g^H_{x^I \Upsilon}}{\partial u^K} \;,
\]
we obtain, after some rearrangements, an expression for the
antiholomorphic derivative
\[
\frac{\partial}{\partial \bar{Y}^I} g^H_{\Upsilon \Upsilon}  =
\left. \frac{\partial g^H_{x^I \Upsilon}}{\partial \Upsilon} \right|_{u} +
\frac{\partial u^K}{\partial \Upsilon} \frac{\partial g^H_{x^I \Upsilon}}{\partial
u^K} - \frac{\partial g^H_{\Upsilon \Upsilon}}{\partial Y^I} -
\frac{\partial u^K}{\partial x^I} \frac{\partial g^H_{\Upsilon \Upsilon}}{\partial
u^K} \;.
\]
After a lengthy but straightforward calculation similar to the one in subsection \ref{hessianstr1}, we find 
\begin{eqnarray}
\label{anomhess}
- i \frac{\partial}{\partial \bar{Y}^I} D_\Upsilon F_{\Upsilon} &=&
\frac{\partial g^H_{\Upsilon \Upsilon}}{\partial \bar{Y}^I} \\
&=&
i \bar{F}_{\bar{I} \Upsilon \Upsilon} + 
2 ( F_{S\Upsilon} + \bar{F}_{\bar{S} \Upsilon}) N_-^{SL}
\left[ \bar{F}_{\bar{I}\bar{L} \Upsilon} - \bar{F}_{\bar{I}L \Upsilon} \right]
\nonumber\\
&& -
i (F_{S\Upsilon} + \bar{F}_{\bar{S}\Upsilon}) N_-^{SL} 
( F_{Q\Upsilon} + \bar{F}_{\bar{Q} \Upsilon}) N_-^{QP}
\left[
\bar{F}_{\bar{I} \bar{P} \bar{L}} + \bar{F}_{\bar{I} PL} - 2 {\bar F}_{{\bar I} {\bar P} L} 
\right] \:. \nonumber
\end{eqnarray}
In the holomorphic case, this reduces to \eqref{HA2}.
In the non-holomorphic case based on (\ref{def-beta1})
it can be readily verified that when setting $\Upsilon=0$, one obtains the holomorphic anomaly equation \eqref{fullhaeg2} for $F^{(2)}$
to leading order in $\beta$ and $N^{-1}_{(0)}$. Namely, using \eqref{def-beta1} and setting $\Upsilon =0$,
\eqref{anomhess} reduces to 
\begin{eqnarray}
\left.  \frac{\partial}{\partial \bar{Y}^I} D_\Upsilon F_{\Upsilon}\right|_{\Upsilon =0} = 2 \frac{\partial}{\partial
\bar{\cal Y}^I} F^{(2)}
= \left. \left( - \bar{F}_{\bar{I} \Upsilon \Upsilon} 
+
 F_{S\Upsilon}  N_-^{SL} 
F_{Q\Upsilon} N_-^{QP}
\bar{F}_{\bar{I} \bar{P} \bar{L}} \right) \right|_{\Upsilon=0} \;. \nonumber
\end{eqnarray}
Using ${\bar F}_{\bar I \Upsilon \Upsilon}|_{\Upsilon =0} = - 2i \Omega_{
\bar I \Upsilon \Upsilon}|_{\Upsilon =0} = - 8 \beta {\bar F}_{\bar I}^{(0) KL} f^{(1)}_{KL}
= 4i \beta {\bar F}_{\bar I}^{(0) KL} F^{(1)}_{KL}$, and redefining $F^{(n)} \rightarrow 2i F^{(n)}$, we obtain
\eqref{fullhaeg2} for $F^{(2)}$.

We note that, in principle, one may now proceed to derive the holomorphic anomaly equation for the higher $F^{(n)}$ (with $n \geq 3$)
by applying covariant derivatives $D_{\Upsilon}$ to \eqref{anomhess}, and subsequently setting $\Upsilon=0$,
as in subsection \ref{hessianstr1}.

We finish by verifying that 
the component $S_{x^I \Upsilon \bar \Upsilon}
= \partial_{x^I} g^H_{\Upsilon \bar \Upsilon}$,
which is constructed out of the 
metric component
$g^H_{\Upsilon \bar \Upsilon} = - i D_{\Upsilon} F_{\bar \Upsilon}$,
does not give rise to an additional non-trivial
differential equation.
Evaluating the relation 
$S_{x^I \Upsilon  \bar \Upsilon} = S_{\bar \Upsilon x^I \Upsilon} = \partial_{\bar \Upsilon} \,  g^H_{x^I \Upsilon}$
in supergravity variables we find that it is identically satisfied. Thus, the only non-trivial differential equation resulting from 
$g^H_{\Upsilon \Upsilon}$ and $g^H_{\Upsilon \bar \Upsilon}$ is encoded in the relation $S_{x^I \Upsilon \Upsilon} = S_{\Upsilon x^I \Upsilon}$.

\section{Concluding remarks \label{sec:concl}}

Let us conclude with a comparison of the 
approach taken in \cite{Cardoso:2014kwa} and the one taken here for obtaining the holomorphic anomaly equation. 
Both are based on the Hesse potential, which is obtained by a Legendre
transform, see \eqref{lforh}.  In the approach of \cite{Cardoso:2014kwa} one works directly
with the Hesse potential, while here we work with the 
associated Hessian structure ($g^H, \nabla$) on the extended 
scalar manifold $\hat{M} = M \times \mathbbm{C}$.
The Hesse potential, which is akin to a Hamiltonian \cite{Cardoso:2014kwa}, is a symplectic function of the special real coordinates.
It can also be expressed either in terms of supergravity variables $(Y^I, \Upsilon)$ (and their complex conjugate) or in terms of stringy (or covariant)
variables $({\cal Y}^I, \Upsilon)$  (and their complex conjugate), see \eqref{oldnewc}.
The approach in \cite{Cardoso:2014kwa} consisted in first expressing the Hesse potential in terms of covariant variables
by means of a power series expansion in $\Delta Y^I$, and  then expressing $\Delta Y^I$
in terms of a power series 
in derivatives of $\Omega$, which we recall was introduced in \eqref{gF=F+Omega}. In this way it was shown in \cite{Cardoso:2014kwa} that the Hesse potential, when
 expressed in terms of covariant variables, equals an infinite sum of symplectic functions, which were denoted
 by ${\cal H}^{(a)}_i$. 
 The label $a$ indicates that the leading term is of order $\Omega^a$.
 For higher values of $a$ there are several functions (labelled by $i =1,2, \dots$) with the same value of $a$.
This decomposition is
 unique. In this decomposition there is only one function, namely ${\cal H}^{(1)}$,
 whose leading term is $\Omega$ itself, while the leading term of 
 all the other ${\cal H}^{(a)}_i$ (with $a \geq 2$) involves derivatives of $\Omega$. 
 In  \cite{Cardoso:2014kwa} it was shown that 
 ${\cal H}^{(1)}$ comprises a subsector of the full Hesse potential that encodes the holomorphic anomaly equation.
 This was achieved by using the explicit expression for ${\cal H}^{(1)}$, which consists of an infinite sum that
 starts with 
  $4 \Omega$, and that involves terms of higher and higher
 powers of derivatives of $\Omega$. By using the fact that $\Omega$ depends on $\Upsilon$, this
 infinite sum was in turn rewritten as a series expansion in $\Upsilon$, with coefficient
 functions that are again symplectic functions.  When $\Omega$ is taken to be harmonic, these symplectic functions, denoted by
 $F^{(n)}$  (the first three of which we display in \eqref{expf123}), satisfy the holomorphic anomaly equation 
\eqref{anomal} with $\alpha =0$. Subsequently, by deforming $\Omega$ by $\alpha$-dependent terms, as in \eqref{omalbe},
it was shown \cite{Cardoso:2014kwa} that the resulting functions $F^{(n)}$ 
  satisfy the full holomorphic anomaly equation 
\eqref{anomal}. 

Thus, summarizing, it was shown in \cite{Cardoso:2014kwa} that the full Hesse potential, when expressed in terms of covariant
variables, 
contains a subsector ${\cal H}^{(1)}$ that, in turn, contains an infinite set of functions $F^{(n)}$   that 
 satisfy the full holomorphic anomaly equation 
\eqref{anomal}.  The other sectors, described by the other functions ${\cal H}^{(a)}_i$, are constructed out of derivatives of $\Omega$,
and thus contain derived information. They are nevertheless important, since they are needed to build up the full Hesse potential.

 In the approach taken in this paper, we instead work with the Hessian metric $g^H$ associated with the full Hesse potential. We work in
 supergravity variables, and we  focus on particular components of $g^H$, namely on $g^H_{\Upsilon \Upsilon}$ and $g^H_{\Upsilon \bar \Upsilon}$.
 These two components (both of which are given in terms of the symplectic covariant derivative introduced in \cite{deWit:1996ix,Cardoso:2012nh})  encode
 different information.
  For instance, when evaluated at $\Upsilon =0$, $g^H_{\Upsilon \Upsilon}$ gives the symplectic function $F^{(2)}$, while 
 $g^H_{\Upsilon \bar \Upsilon}$ gives the symplectic function ${\cal H}^{(2)}$ evaluated at $\Upsilon =0$. This is reminiscent of the decomposition 
 of the full Hesse potential into symplectic functions ${\cal H}^{(a)}_i$ discussed above. We then consider the totally symmetric rank three tensor $S = \nabla g^H$,
 and we first focus on its component $S_{x^I \Upsilon \Upsilon} = \partial _{x^I } \, g^H_{\Upsilon \Upsilon}$. Evaluating the relation 
 $S_{x^I \Upsilon \Upsilon} = S_{\Upsilon x^I \Upsilon}$ in supergravity variables and subsequently setting $\Upsilon =0$ we obtain the holomorphic
 anomaly equation for $F^{(2)}$. One may then ask whether other components of $S$ will lead to additional non-trivial differential equations.
 To address this, we consider the component $S_{x^I \Upsilon  \bar \Upsilon} = \partial _{x^I } \, g^H_{\Upsilon \bar  \Upsilon}$.
   Evaluating the relation 
 $S_{x^I \Upsilon  \bar \Upsilon} = S_{\bar \Upsilon x^I \Upsilon} = \partial_{\bar \Upsilon} \,  g^H_{x^I \Upsilon}$
 in supergravity variables we find that it is identically satisfied. Thus, we conclude that the only non-trivial differential equation resulting from 
 $g^H_{\Upsilon \Upsilon}$ and $g^H_{\Upsilon \bar \Upsilon}$ is encoded in 
 the relation $S_{x^I \Upsilon \Upsilon} = S_{\Upsilon x^I \Upsilon}$.

\subsection*{Acknowledgements}

\noindent
We would like to thank Vicente Cort\'es and Bernard
de Wit for valuable discussions.
The
work of G.L.C. was partially
supported by FCT/Portugal through \\UID/MAT/04459/2013 and through
EXCL/MAT-GEO/0222/2012. 
The work of T.M. was, in part, supported by STFC and by
a BCC visiting fellowship associated to the excellency grant EXCL/MAT-GEO/0222/2012.
This work was also supported by the COST
action MP1210 {\it ``The String Theory Universe''}. \\
T.M. would like to thank CAMGSD (Department of Mathematics, IST) for the BCC fellowship which enabled him
to visit IST, and for the kind hospitality. 
G.L.C. would like to thank the Max-Planck-Institut f\"ur Gravitationsphysik (Albert-Einstein-Institute) for kind hospitality during the completion of this work.

\begin{appendix}
\section{Connections on  vector bundles \label{diff_geo}}

We review some standard facts about connections in 
vector bundles. Let $E\rightarrow M$ be a vector bundle over
a manifold $M$. Then a connection or 
covariant derivative $\nabla$ on $E$ is a map 
\[
\nabla \,: \mathfrak{X}(M) \times \Gamma(E) \rightarrow \Gamma(E) \;,\;\;
(X, s) \mapsto \nabla_X s
\]
which is a linear derivation (satisfies the product rule) with
respect to sections $s \in \Gamma(E)$ of $E$,  and which 
is $C^\infty(M)$-linear with respect to vector fields
in $X \in \mathfrak{X}(M)$.  

For vector bundles of the form $\Omega^p(M,E)=
\Lambda^p T^* M \otimes E$, that is
for bundles of $p$-forms with values in a vector bundle $E$, the 
covariant exterior derivative 
\[
d_\nabla \;: \Omega^p(M,E) \rightarrow \Omega^{p+1}(M,E)
\] is uniquely determined by
its action on sections of $E$. For a given basis $\{ s_a \}$ 
of sections one sets
\[
d_\nabla s_a = \nabla s_a = 
\omega_a^b s_b \;,
\]
where $\omega_a^b$ is the connection one-form of $\nabla$. The
derivative of a general section $s=f^a s_a \in \Omega^0(M,E)$, 
where $f^a \in C^\infty(M)$ is then
\[
d_\nabla s = df^a \otimes s_a + f^a \omega^b_a s_b \;.
\]
The extension to forms of degree $p>0$ is completely determined
by linearity and the product rule
\[
d_\nabla (\alpha \otimes s) = d\alpha \otimes s + (-1)^{\deg \alpha}
\alpha \wedge d_\nabla s \;,
\]
where $\alpha \in \Omega^p(M) $.
The exterior covariant derivative $d_\nabla$  of $\omega \in \Omega^p(M,E)$ can be expressed 
in terms of the covariant derivative $\nabla$ by
\begin{eqnarray*}
(d_\nabla \omega) (X_0, \ldots , X_p) &=&
\sum_{l=0}^p (-1)^l \nabla_{X_l} ( \omega( \ldots \hat X_l \ldots )) \\
&& +
\sum_{i<j} (-1)^{i+j} \omega( [X_i, X_j], \ldots \hat{X}_i \ldots \hat{X}_j 
\ldots ) \;,
\end{eqnarray*}
where  $X_0, \ldots X_p$ are vector fields, and where 
$\hat{X}$ indicates that the corresponding vector field 
is omitted.

The curvature of the connection $\nabla$ is given by
$R^\nabla (s) = d_\nabla (d_\nabla s)$. If the connection $\nabla$
is flat, then $d_\nabla^2=0$, so that $d_\nabla$ defines an exact
sequence. In this case a version of the Poincar\'e lemma exists.

\section{Special coordinates \label{App:special_coord}}

One of the defining conditions of affine special K\"ahler geometry
is $d_\nabla J=0$, where $J$ is the complex structure. We can apply
the above results since $J$ is a section of $\mbox{End}(E) \simeq
E^* \otimes E$, where $E=TM$. In the following we derive the
local form (\ref{curlJ}) of (\ref{dNablaJ}), and also explain
how the existence of special coordinates can be derived.

Let $A$ be a section of $\mbox{End}(TM)$. 
Choosing
dual bases $\{ e^a \}, \{ e_a \}$ of sections, and regarding
$A$ as a $TM$-valued one-form, we have
\[
A = A^a e_a = A^a_{\;\;b} e^b \otimes e_a \in \Omega^1(M,TM)
\]
and
\[
d_\nabla A = dA^a \otimes e_a - A^a \wedge \omega_a^{\;\;b} e_b
\in \Omega^2(M,TM) \;.
\]
If the connection $\nabla$ is flat, we can choose sections
$e_a$ such that the connection one-form vanishes, $\omega_a^{\;\;b}=0$
and in such a frame $d_\nabla A=0$ reduces to $dA^a=0$. Thus
the one-forms $A^a$ are locally exact, $A^a=d\phi^a$. 

Since $E=TM$, the torsion of the connection $\nabla$ is defined
by  
\[
T^\nabla(X,Y) = \nabla_X Y - \nabla_Y X - [X,Y] \;.
\]
One can show that 
\[
T^\nabla(X,Y) = d_\nabla \mbox{Id}(X,Y) \;,
\]
where 
\[
\mbox{Id} = e^a \otimes e_a = \delta^b_a e^a \otimes e_b 
\] 
is the identity endomorpism of $TM$. 
If $\nabla$ is both flat and torsion free, then $d_\nabla \mbox{Id}=0$
implies that in a frame where the connection vanishes, the
one-forms $I^a= e^a$ are locally exact $e^a = dt^a$. This defines
a set of $\nabla$-affine coordinates. In such coordinates 
the condition (\ref{dNablaJ}) becomes
\[
d_\nabla J = 0  \Rightarrow d J^a = 0 \Rightarrow \partial_{[b} J^b_{\;\;c]}
=0  \;.
\]

If the manifold $M$ is in addition equipped with a non-degenerate,
closed two-form $\omega \in \Omega^2(M)$, $d\omega=0$, then a connection
$\nabla$ is called symplectic if the symplectic form $\omega$ is
parallel:
\[
\nabla \omega = 0 \Leftrightarrow \nabla_X \omega =
X(\omega_{ab}) e^a \wedge e^b +
\omega_{ab} (\nabla_X e^a) \wedge e^b +
\omega_{ab} e^a \wedge \nabla_X e^b =0 \;,
\]
for all vector fields $X$. 
If the connection $\nabla$ is in addition flat, we can choose sections
$e^a$ such that $\nabla_X e^a=0$. With respect to such a basis
the coefficients of $\omega$ are constant, $X(\omega_{ab})=0$. 
If the connection $\nabla$ is in addition torsion-free, the co-frame
$e^a$ comes from an affine coordinate system $t^a$, and 
\[
\omega = \frac{1}{2} \omega_{ab} dt^a \wedge dt^b
\]
where $\omega_{ab}$ is a constant, antisymmetric, non-degenerate matrix. 
Using the linear part of the affine transformation that we can apply to $t^a$, 
the matrix $\omega_{ab}$ can be brought to the standard form
\[
(\omega_{ab})^{\rm Standard} = \left( \begin{array}{cc}
0 & \mathbbm{1} \\
-\mathbbm{1} & 0 \\
\end{array} \right) \;.
\]
This form is still invariant under affine transformations where the
linear part is symplectic. The associated coordinates are called
Darboux coordinates.  Thus we have seen that for a flat, torsion-free,
symplectic connection the $\nabla$-affine coordinates can be chosen
to be Darboux coordinates. In the context of affine special
K\"ahler geometry such coordinates are called special real coordinates.

We remark that in the main part of this paper we use
special real coordinates $(q^a) = (x^I,y_I)$ where
the K\"ahler form takes the form
\[
\omega = 2 dx^I \wedge dy_I = \Omega_{ab} dq^a \wedge dq^b  \;.
\]
Note that the components of $\omega$ with respect to the coordinates
$(q^a)$ are 
$\omega_{ab} = 2 \Omega_{ab} = 2 \omega_{ab}^{\rm Standard}$. In other
words the special real coordinates differ from standard Darboux 
coordinates by a conventional factor $\sqrt{2}$.

\section{Symplectic transformations and functions \label{symtr}}

Symplectic transformations acts as follows on a symplectic vector $(Y^I, F_I)$ ($I=0, \dots n$),
\begin{eqnarray}
Y^I & \rightarrow & U^I{}_J Y^J + Z^{IJ} F_J \;, \nonumber\\
F_I &   \rightarrow & V_I{}^J F_J + W_{IJ} Y^J \;, \nonumber
\end{eqnarray}
where $U, V, Z, W$ are the $(n+1) \times (n+1)$ real submatrices that give rise to an element of
$Sp(2n+2, \mathbb{R})$. When $F$ equals the prepotential $F^{(0)}$, then $N^{IJ}_{(0)}$ 
transforms as follows under symplectic transformations \cite{deWit:1996ix}, 
\begin{equation}
N^{IJ}_{(0)} \rightarrow \bar{\cal S}_0{}^I{}_K \, {\cal S}_0{}^J{}_L N_{(0)}^{KL} =
{\cal S}_0{}^I{}_K \, {\cal S}_0{}^J{}_L \left( N_{(0)}^{KL} - i {\cal Z}_0^{KL} \right) \;,
\label{transfN}
\end{equation}
where
\begin{eqnarray}
{\cal S}_0{}^I{}_K &=& U^I{}_K  + Z^{IJ} F^{(0)}_{JK} \;, \nonumber\\
{\cal Z}_0^{IJ} &=& [{\cal S}^{-1}_0]^I{}_K Z^{KJ} \;.
\label{SZ}
\end{eqnarray}

Consider a generalized prepotential $F(Y, \bar Y, \Upsilon, \bar \Upsilon) = F^{(0)}(Y) + 2 i \Omega (Y, \bar Y, \Upsilon, \bar \Upsilon)$, where 
$\Omega$ is taken to be harmonic, $\Omega (Y, \bar Y, \Upsilon, \bar \Upsilon) = f(Y, \Upsilon) + \bar f (\bar Y, \bar \Upsilon)$. 
Expanding $f(Y, \Upsilon)$ as in \eqref{serom} and inserting this into \eqref{Fnhol}, yields explicit expressions for
the symplectic functions $F^{(n)}$. The first three read as follows,
\begin{eqnarray}
\label{expf123}
F^{(1)} &=& 2i \, f^{(1)} \;, \\
F^{(2)} &=& 2 i \left( f^{(2)} - N^{IJ}_{(0)} \, f^{(1)}_I f^{(1)}_J \right)\;,  \nonumber\\
F^{(3)} &=& 2i \left(  f^{(3)} - 2 N^{IJ}_{(0)} \, f^{(2)}_I f^{(1)}_J + 2 f^{(1)}_I N^{IJ}_{(0)} f^{(1)}_{JK} N^{KL}_{(0)}
f^{(1)}_L \right. \nonumber \\
 & & 
\left.  + \frac{2i}{3} F^{(0)}_{IJK} N^{IP}_{(0)}N^{JQ}_{(0)}N^{KR}_{(0)}
f_P^{(1)} f_Q^{(1)} f_R^{(1)} 
\right) \;,\nonumber
\end{eqnarray}
in accordance with \cite{Cardoso:2014kwa}. These expressions get modified when $\Omega$ is not any longer harmonic.
The resulting expressions for the topological string can be found in appendix D of \cite{Cardoso:2014kwa}.

\end{appendix}

\providecommand{\href}[2]{#2}\begingroup\raggedright\endgroup

\end{document}